\documentclass[english,american,aps,rmp,reprint,superscriptaddress]{revtex4-1}
\usepackage[T1]{fontenc}
\usepackage[latin9]{inputenc}
\usepackage{color}
\usepackage{babel}
\usepackage{array}
\usepackage{multirow}
\usepackage{amsmath}
\usepackage{amssymb}
\usepackage{graphicx}
\usepackage[unicode=true,pdfusetitle,
 bookmarks=true,bookmarksnumbered=false,bookmarksopen=false,
 breaklinks=false,pdfborder={0 0 0},backref=false,colorlinks=true]
 {hyperref}
\hypersetup{
 colorlinks,linkcolor=myurlcolor,citecolor=myurlcolor,urlcolor=myurlcolor}

\makeatletter

\providecommand{\tabularnewline}{\\}

\newenvironment{lyxlist}[1]
{\begin{list}{}
{\settowidth{\labelwidth}{#1}
 \setlength{\leftmargin}{\labelwidth}
 \addtolength{\leftmargin}{\labelsep}
 }}
{\end{list}}

\usepackage{color}
\usepackage{babel}
\usepackage{array}
\usepackage{txfonts}
\usepackage{braket}
\usepackage{bbm}

\usepackage{colortbl}
\definecolor{myurlcolor}{rgb}{0.75,0,0.75}

\makeatother

\begin{document}

\title{\emph{Colloquium}: Quantum Coherence as a Resource}

\author{Alexander Streltsov}

\email{streltsov.physics@gmail.com}

\affiliation{$\mbox{Faculty of Applied Physics and Mathematics, Gda\'{n}sk University of Technology, 80-233 Gda\'{n}sk, Poland}$}

\affiliation{$\mbox{National Quantum Information Centre in Gda\'{n}sk, 81-824 Sopot, Poland}$}

\affiliation{$\mbox{Dahlem Center for Complex Quantum Systems, Freie Universität Berlin, D-14195 Berlin, Germany}$}

\affiliation{$\mbox{ICFO -- Institut de Ciències Fotòniques, The Barcelona Institute of Science and Technology, ES-08860 Castelldefels, Spain}$}

\author{Gerardo Adesso}

\email{gerardo.adesso@nottingham.ac.uk}

\affiliation{$\mbox{Centre for the Mathematics and Theoretical Physics of Quantum Non-Equilibrium Systems (CQNE),}$
\\
 $\mbox{School of Mathematical Sciences, The University of Nottingham, University Park, Nottingham NG7 2RD, United Kingdom}$}

\author{Martin B. Plenio}

\email{martin.plenio@uni-ulm.de}

\affiliation{$\mbox{Institute of Theoretical Physics \& IQST, University of Ulm, Albert-Einstein-Allee 11, D-89069 Ulm, Germany}$}

\date{July 10, 2017}
\begin{abstract}
The coherent superposition of states, in combination with the quantization of observables,
represents one of the most fundamental features that mark the departure
of quantum mechanics from the classical realm. Quantum coherence in
many-body systems embodies the essence of entanglement and is an essential
ingredient for a plethora of physical phenomena in quantum optics,
quantum information, solid state physics, and nanoscale thermodynamics.
In recent years, research on the presence and functional role of quantum
coherence in biological systems has also attracted a considerable
interest. Despite the fundamental importance of quantum coherence,
the development of a rigorous theory of quantum coherence as a physical
resource has only been initiated recently. In this Colloquium we discuss
and review the development of this rapidly growing research field
that encompasses the characterization, quantification, manipulation,
dynamical evolution, and operational application of quantum coherence.
\end{abstract}
\maketitle
\tableofcontents{}

\clearpage

\section{Introduction}

\label{sec:introduction}

Coherence marks the departure of today's theories of the physical
world from the principles of classical physics. The theory of electro-magnetic
waves, which may exhibit optical coherence and interference, represents
a radical departure from classical ray optics. Energy quantisation
and the rise of quantum mechanics as a unified picture of waves and
particles in the early part of the 20$^{\text{th}}$ century has further
amplified the prominent role of coherence in physics. Indeed, by combination
of energy quantization and the tensor product structure of the state
space, coherence underlies phenomena such as quantum interference
and multipartite entanglement, that play a central role in applications of quantum
physics and quantum information science.

The investigation and exploitation of coherence of quantum \textit{optical
fields} has a longstanding history. It has enabled the realization
of now mature technologies, such as the laser and its applications,
that are often classified as `Quantum Technologies 1.0' as they rely
mainly on single particle coherence. At the mathematical level the
coherence of quantum optical fields is described in terms of phase
space distributions and multi-point correlation functions, approaches
that find their roots in classical electromagnetic theory \cite{Sudarshan1963,Glauber1963,Mandel1965}.

However, quantum coherence is not restricted to optical fields. More
importantly, as the key ingredient that drives quantum technologies,
it would be highly desirable to be able to precisely quantify the
usefulness of \textit{coherence as a resource} for such applications.
These pressing questions are calling for a further development of
the theory of quantum coherence.

The emergence of quantum information science over the last three decades
has, amongst other insights, led to a reassessment of quantum physical
phenomena as resources that may be exploited to achieve tasks that
are otherwise not possible within the realm of classical physics.
This resource-driven viewpoint has motivated the development of a
quantitative theory that captures the resource character of physical
traits in a mathematically rigorous fashion.

In a nutshell, any such theory first considers constraints that are
imposed on us in a specific physical situation (e.g.~the inability
to perform joint quantum operations between distant laboratories due
the impossibility to transfer quantum systems from one location to the other while preserving their quantum coherence, and thus restricting us to local operations and classical communication).
Executing general quantum operations under such a constraint then
requires quantum states that contain a relevant resource (e.g.~entangled
states that are provided to us at a certain cost) and can be consumed
in the process. The formulation of such resource theories was in fact
initially pursued with the quantitative theory of entanglement \cite{Plenio2007,HorodeckiRMP09}
but has since spread to encompass a wider range of operational settings
\cite{HorodeckiO2013,coecke2014mathematical,resourcetheoriesofknowledge}.

The theory of quantum coherence as a resource is a case in point.
Following an early approach to quantifying {\it superpositions} of orthogonal quantum states by \cite{Aberg2006}, and progressing alongside the independent yet related resource theory of {\it asymmetry} \cite{Vaccaro2008,Gour2008,Gour2009,Marvian2014a,Marvian2014}, a {\it resource theory of coherence} has been primarily proposed in  \cite{Baumgratz2014,Levi14} and further developed in \cite{Winter2015,Yadin2015b,Chitambar2016,Chitambar2016b,Chitambar2017}. Such a theory asks the question what can be achieved and
at what resource cost when the devices that are available to us are
essentially classical, that is, they cannot create coherence in a
preferred basis. This analysis, currently still under development,
endeavors to provide a rigorous framework to describe quantum
coherence, in analogy with what has been done for quantum entanglement
and other nonclassical resources \cite{Plenio2007,HorodeckiO2013,HorodeckiRMP09,Modi2012,ABC,Streltsov2014,VogelScr}.
Within such a framework, recent progress has shown that a growing number of applications can be certified
to rely on various incarnations of quantum coherence as a primary ingredient, and appropriate
figures of merit for such applications can be precisely linked back
to specific coherence monotones, providing operational interpretations
for the latter.

These applications include so-called `Quantum Technologies
2.0', such as quantum-enhanced metrology and communication protocols,
and extend further into other fields, like thermodynamics and even
certain branches of biology. Beyond such application-driven viewpoint, which may provide new insights into all these areas,
one can also consider the theory of coherence as a resource as a novel
approach towards the demarcation of the fundamental difference between classical
and quantum physics in a quantitative manner: a goal that may eventually
lead to a better understanding of the classical-quantum boundary.

\begin{figure}[t]
\centering \includegraphics[width=8.5cm]{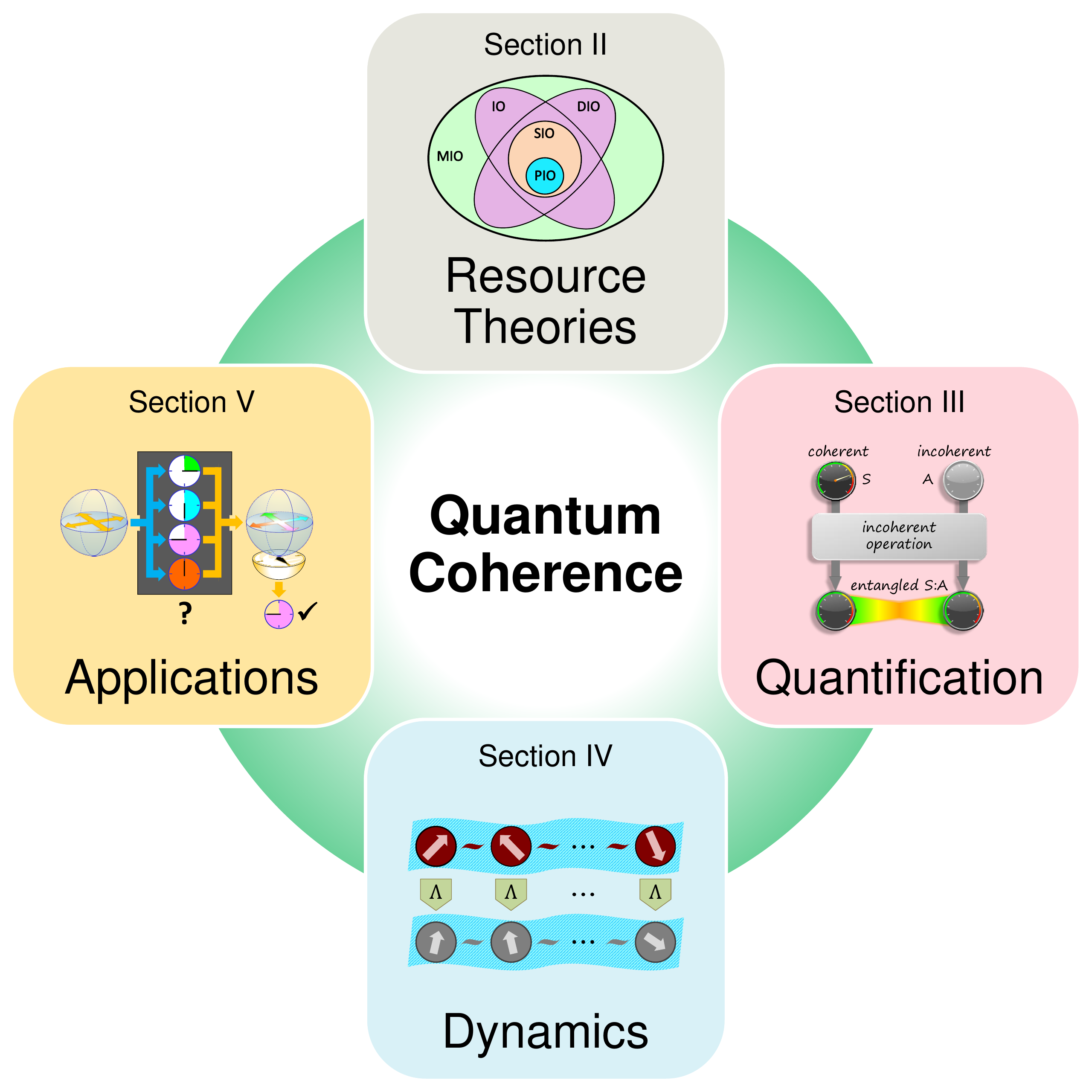} \caption{(Color online) Plan of the Colloquium. (Top) Section~\ref{sec:resource-theories}: \textit{Resource
theories of quantum coherence}; the inset depicts a comparison between
some classes of incoherent operations, adapted and reproduced with permission from \cite{Chitambar2016}. (Right)
Section~\ref{sec:Quantifying-coherence}: \textit{Quantifying quantum
coherence}; the inset depicts the construction of coherence monotones
from entanglement, adapted and reproduced with permission from \cite{Streltsov2015b}. (Bottom) Section~\ref{sec:dynamics}:
\textit{Dynamics of quantum coherence}; the inset depicts an illustration
of coherence freezing under local incoherent channels, adapted and reproduced with permission from
\cite{Bromley2014}. (Left) Section~\ref{sec:applications}: \textit{Applications
of quantum coherence}; the insets depicts a schematic of a quantum
phase discrimination protocol, adapted and reproduced with permission from \cite{Napoli2016}. \textit{Introduction}
(Section~\ref{sec:introduction}) and \textit{Conclusions} (Section~\ref{sec:conclusions})
complete the Colloquium. \label{fig:plan}}
\end{figure}

The present Colloquium collects the most up-to-date knowledge on coherence
in single and composite quantum systems, from a modern information
theory perspective. We set to review this fascinating and fundamental
subject in an accessible manner, yet without compromising any rigor.

The Colloquium is organized as follows (see Figure~\ref{fig:plan}).
Section~\ref{sec:resource-theories} gives a comprehensive overview
of recent developments to construct a resource theory of quantum coherence,
including a hierarchy of possible classes of incoherent operations,
and the conditions to which any valid coherence quantifier should
abide; it also discusses established links with other resource theories,
most prominently those of asymmetry and of quantum entanglement. Section~\ref{sec:Quantifying-coherence}
presents a compendium of recently proposed monotones and measures
of quantum coherence, based on different physical approaches endowed
with different mathematical properties, in single and multipartite
systems; interplays with other measures of nonclassicality are highlighted
as well. Section~\ref{sec:dynamics} reviews the phenomenology of
quantum coherence in the dynamical evolution of open quantum systems,
further reporting results on the average coherence of random states,
and on the cohering power of quantum channels. Section~\ref{sec:applications}
focuses on the plethora of applications and operational interpretations
highlighted so far for quantum coherence  in thermodynamics, interference phenomena, quantum
algorithms, metrology and discrimination, quantum biology, many-body
physics and the detection of quantum correlations. Section~\ref{sec:conclusions}
concludes the Colloquium with a summary and a discussion of some currently
open issues in the theoretical description of coherence and its role
in quantum physics and beyond.

We emphasize that, due to limitations in space and focus, this Colloquium
cannot cover all ramifications of the concept of quantum coherence.
It is nevertheless our expectation that this
Colloquium, while being self-contained, can stimulate the reader to
undertake further research towards achieving a fully satisfactory
and physically consistent characterization of the wide-interest topic
of coherence as a resource in quantum systems of arbitrary dimensions.
This, we hope, may also lead to the formulation of novel direct applications
of coherence (or an optimization of existing ones) in a variety of
physical and biological contexts of high technological interest.

\section{\label{sec:resource-theories}Resource theories of quantum coherence}

Coherence is a property of the physical world that is used to drive
a wide variety of phenomena and devices. Hence coherence adopts the
quality of a resource, as it may be provided at a certain cost, manipulated
by otherwise incoherent means, and consumed to achieve useful tasks.
The quantitative study of these processes and their attainable efficiencies
requires careful definitions of the accessible operations and gives
rise to a framework that has become known as a \textit{resource theory}.
In line with earlier developments in quantum information science,
for example in the context of entanglement \cite{Vedral1998,Plenio2007,Brandao2008,HorodeckiRMP09,Brandao2010},
quantum thermodynamics \cite{Ruch1975,Janzing00,FGSL13,Gour2013,Goold2016},
purity \cite{HorodeckiHO2003}, and reference frames \cite{Gour2008,Gour2009,Marvian2014a},
the formulation of the resource theory of coherence
\cite{Aberg2006,Baumgratz2014,Levi14,Winter2015} extends the family of
resources theories of knowledge \cite{resourcetheoriesofknowledge}.

\subsection{Constraints, operations and resources}

A resource theory is fundamentally determined by constraints that
are imposed on us and which determine the set of the freely accessible
quantum operations ${\cal F}$. These constraints may be due to either
fundamental conservation laws, such as superselection rules and energy
conservation, or constraints due to the practical difficulty of executing
certain operations, e.g.~the restriction to local operations and
classical communication (LOCC) which gives rise to the resource theory
of entanglement \cite{Plenio2007,HorodeckiRMP09}.

The states that can be generated from the maximally mixed state\footnote{The maximally mixed state can always be obtained by erasing all information
about the system. Hence it is fair to assume that it is devoid of
any useful resource and freely available.} by the application of free operations in ${\cal F}$ alone, are considered
to be available free of charge, forming the set ${\cal I}$ of free
states. All the other states attain the status of a resource, whose
provision carries a cost. These resource states may be used to achieve
operations that cannot be realized by using only members of ${\cal F}$.
Alternatively, one may also begin by defining the set of free states
${\cal I}$, and then consider classes of operations that map this
set into itself and use this to define ${\cal F}$. For the purposes
of the present exposition of the resource theory of coherence, we
begin by adopting the latter point of view and then proceed to require
additional desirable properties of our classes of free operations.

\subsubsection{Incoherent states}

Coherence is naturally a basis dependent concept, which is why we first
need to fix the preferred, or \textit{reference basis} in which to
formulate our resource theory.
% \ga{FOOTNOTE REMOVED}
%\footnote{This is analogous to, say,
%entanglement theory where a global basis change will alter our notion of
%entanglement. Nevertheless, entanglement is invariant under a change of
%the local bases.}.
The reference basis may be dictated by the physics of the problem
under investigation (e.g.~one may focus on the energy eigenbasis
when addressing coherence in transport phenomena and thermodynamics)
or by a task for which coherence is required (e.g.~the estimation
of a magnetic field in a certain direction within a quantum metrology
setting). Given a $d$-dimensional Hilbert space $\mathcal{H}$ (with
$d$ assumed finite, even though some extensions to infinite $d$
can be considered), we will denote its reference orthonormal basis
by $\{|i\rangle\}_{i=0,\ldots,d-1}$. The density matrices that are
diagonal in this specific basis are called \textit{incoherent}, i.e.,~they
are accessible free of charge, and form the set $\mathcal{I}\subset\mathcal{B}(\mathcal{H})$,
where ${\cal B}({\cal H})$ denotes the set of all bounded trace class
operators on ${\cal H}$. Hence, all incoherent density operators
$\varrho\in\mathcal{I}$ are of the form
\begin{equation}
\varrho=\sum_{i=0}^{d-1}p_{i}|i\rangle\langle i|\label{eq:incoherent}
\end{equation}
with probabilities $p_{i}$.

In the case of more than one party, the preferred basis with respect
to which coherence is studied will be constructed as the tensor product
of the corresponding local reference basis states for each subsystem.
General multipartite incoherent states are then defined as convex
combinations of such incoherent pure product states \cite{Bromley2014,Streltsov2015b,Winter2015}.

For example, if the reference basis for a single qubit is taken to
be the computational basis $\{\ket{0},\ket{1}\}$, i.e., the eigenbasis
of the Pauli $\sigma_{z}$ operator, then any density matrix with
a nonzero offdiagonal element $|\varrho_{01}|=|\!\braket{0|\varrho|1}\!|\neq0$
is outside the set ${\cal I}$ of incoherent states, hence has a resource
content. Similarly, for an $N$-qubit system, the set of incoherent
states ${\cal I}$ is formed by all and only the density matrices
$\varrho$ diagonal in the composite computational basis $\{\ket{0},\ket{1}\}^{\otimes N}$,
with any other state being coherent, that is, resourceful.

We also note that some frameworks of coherence may allow for a
larger set of free states. This is in particular the case for
the resource theory of asymmetry \cite{Gour2008,Gour2009,Marvian2014a,Marvian2014},
where the set of free states is defined by all states which commute
with a given Hamiltonian $H$. If the Hamiltonian is nondegenerate,
the corresponding set of free states is exactly the set of incoherent
states described above, where the incoherent basis is defined by the
eigenbasis of the Hamiltonian. However, the situation changes if the
Hamiltonian has degeneracies, in which case also any superposition
of the eigenstates corresponding to the degenerate subspaces is also
considered as free. This has important implications in quantum thermodynamics,
as is described in more detail in Section~\ref{sec:thermodynamics}.
In the following, whenever we refer to incoherent states, we explicitly
mean states of the form~(\ref{eq:incoherent}).

\subsubsection{\label{sub:Classes}Classes of incoherent operations}

The definition of free operations for the resource theory of coherence is not unique and different
choices, often motivated by suitable practical considerations,
are being examined in the literature. Here, we present the most important
classes and briefly discuss their properties and relations among each
other.

We start with the largest class, the \emph{maximally incoherent operations}
(MIO) \cite{Aberg2006} (also known as incoherence preserving operations),
which are defined as any trace preserving completely positive and
non-selective quantum operations $\Lambda:\mathcal{B}(\mathcal{H})\mapsto\mathcal{B}(\mathcal{H})$
such that
\begin{equation}
\Lambda[\mathcal{I}]\subseteq\mathcal{I}.
\end{equation}
As with every quantum operation, also this mathematically natural
set of operations can always be obtained by a Stinespring dilation,
i.e.~the provision of an ancillary environment in some state $\sigma$, a subsequent
unitary operation $U$ between system and environment, followed by
the tracing out of the environment,
\begin{equation}
\Lambda[\varrho]=\mathrm{Tr}_{E}[U(\varrho\otimes\sigma)U^{\dagger}].\label{eq:Stinespring}
\end{equation}
If an operation can be implemented as in Eq.~(\ref{eq:Stinespring})
by using an incoherent state $\sigma$ of the environment and a global
incoherent unitary $U$ (a unitary that is diagonal in the preferred
basis), we say that the operation has a \emph{free dilation}. It should
be noted that despite the fact that MIO cannot create coherence, these
operations in general do not have a free dilation\footnote{This mirrors the situation in entanglement theory where separable
operations cannot create entanglement but in general cannot be implemented
via LOCC \cite{Bennett1999}.} \cite{Chitambar2016,Chitambar2016b,Chitambar2017,Marvian2016}.

A smaller and more relevant class of free operations for the theory of coherence is that of {\em incoherent
operations} (IO) \cite{Baumgratz2014} which are characterized as
the set of trace preserving completely positive maps $\Lambda:\mathcal{B}(\mathcal{H})\mapsto\mathcal{B}(\mathcal{H})$
admitting a set of \textit{Kraus operators} $\{K_{n}\}$\footnote{According to the Kraus decomposition, the maps act as $\Lambda[\varrho]=\sum_{n}K_{n}\varrho K_{n}^{\dagger}$.}
such that $\sum_{n}K_{n}^{\dagger}K_{n}=\mathbbm{1}$ (trace preservation)
and, for all $n$ and $\varrho\in{\cal I}$,
\begin{eqnarray}
\frac{K_{n}\varrho K_{n}^{\dagger}}{\text{Tr}[K_{n}\varrho K_{n}^{\dagger}]} & \in & {\cal I}.
\end{eqnarray}
This definition of IO ensures that, in any of the possible outcomes
of such an operation, coherence can never be generated from an incoherent
input state, not even probabilistically\footnote{Note that relaxing the
condition of trace preservation here may have non-trivial consequences,
as one should then ensure that the ``missing'' Kraus operators share the
property that they map $\mathcal{I}$ into itself. That this is possible
for IO was recently proven in \cite{Theurer2017}.}. Also this class of
operations does not admit a free dilation in general
\cite{Chitambar2016,Chitambar2016b,Chitambar2017,Marvian2016}.

In the previous two definitions, the focus was placed on the inability
of incoherent operations to generate coherence. One may however be
more stringent by adding further desirable properties to the set of
free operations. One such approach requires that admissible operations
are not capable of making use of coherence in the input state. Defining
the dephasing operation
\begin{equation}
\Delta[\varrho]=\sum_{i=0}^{d-1}|i\rangle\langle i|\varrho|i\rangle\langle i|,\label{eq:dephasing}
\end{equation}
an operation $\Lambda$ is called {\em strictly incoherent} (SIO)
\cite{Winter2015,Yadin2015b} if it can be written in terms of a set
of incoherent Kraus operators $\{K_{n}\}$, such that the outcomes
of a measurement in the reference basis applied to the output are
independent of the coherence of the input state, i.e.,
\begin{equation}
\langle i|K_{n}\varrho K_{n}^{\dagger}|i\rangle=\langle i|K_{n}\Delta[\varrho]K_{n}^{\dagger}|i\rangle
\end{equation}
for all $n$ and $i$. Equivalently, SIO can be characterized as those
operations which have an incoherent Kraus decomposition $\{K_{n}\}$
such that the operators $K_{n}^{\dagger}$ are also incoherent \cite{Winter2015,Yadin2015b}.
As was shown by \cite{Chitambar2016,Chitambar2016b,Chitambar2017}, SIO in general do not admit
a free dilation either. Nevertheless, a special type of dilation
for SIO of the form~(\ref{eq:Stinespring}) was provided in \cite{Yadin2015b}, which consists of: (i) unitary operations on the environment controlled by the incoherent basis of the system: $\sum_j \ket{j}\!\bra{j} \otimes U_j$; (ii) measurements on the environment in any basis; (iii) incoherent unitary operations on the system conditioned on the measurement outcome: {$\sum_j e^{i \theta_j} \ket{\pi(j)}\!\bra{j}$, with $\{\ket{\pi(j)}\}$} denoting a permutation of the reference basis of the system.

The classes of operations defined so far include permutations of the
reference basis states for free. This is natural when any such operation
is considered from a passive point of view, amounting to a relabeling
of the states. Viewed from an active point of view instead, i.e.~asking
for a unitary operation that realizes this permutation, it may be
argued that such an operation does in fact cost coherence resources
in order to be performed in laboratory. This suggests that, from an
operational point of view, permutations should be excluded from the
free operations, as is done in the more stringent set of {\em translationally-invariant
operations} (TIO). The latter are defined as those commuting with
phase randomization \cite{Gour2008,Gour2009,Marvian2013,Marvian2014a,Marvian2014,Marvian2015,Marvian2016}.
More precisely, given a Hamiltonian $H$, an operation $\Lambda$
is translationally invariant with respect to $H$ if it fulfills the
condition \cite{Gour2009,Marvian2014a,Marvian2015}
\begin{equation}
e^{-iHt}\Lambda[\varrho]e^{iHt}=\Lambda[e^{-iHt}\varrho e^{iHt}].\label{transinv}
\end{equation}
TIO play an important role in the resource theory of asymmetry (see
Sec.~\ref{sub:Asymmetry}) and quantum thermodynamics (see Sec.~\ref{sec:thermodynamics}).
Interestingly, \cite{Marvian2016} showed that TIO have a free dilation
if one additionally allows postselection with an incoherent measurement
on the environment.

As already mentioned above, the sets MIO, IO, and SIO in general do
not have a free dilation, i.e., they cannot be implemented by coupling
the system to an environment in an incoherent state followed by a
global incoherent unitary. Motivated by this observation, \cite{Chitambar2016,Chitambar2016b,Chitambar2017}
introduced the set of \emph{physical incoherent operations} (PIO).
These are all operations which can instead be implemented in the aforementioned
way, additionally allowing for incoherent measurements on the environment
and classical postprocessing of the measurement outcomes.

Clearly, MIO is the largest set of free operations for a resource
theory of coherence, and all other sets listed above are strict subsets
of it. Inclusion relations between each of these sets are nontrivial
in general. Here, we mention that (see also
Fig.~\ref{fig:plan}, top panel)
\begin{equation}
\mathrm{PIO}\subset\mathrm{SIO}\subset\mathrm{IO}\subset\mathrm{MIO},
\end{equation}
and refer to \cite{Chitambar2016,Chitambar2016b,Chitambar2017,Streltsov2015d} for a more detailed
discussion.

Another interesting set is given by \emph{dephasing-covariant incoherent
operations} (DIO), which were introduced independently by \cite{Chitambar2016,Chitambar2016b,Chitambar2017}
and \cite{Marvian2016}. These are all operations $\Lambda$ which commute with
the dephasing map Eq.~(\ref{eq:dephasing}), i.e., $\Lambda\big[\Delta[\varrho]\big]=\Delta\big[\Lambda[\varrho]\big]$.
It is an interesting open question whether DIO have a free dilation.

We also mention \emph{genuinely incoherent operations} (GIO) and \emph{fully
incoherent operations} (FIO) \cite{Streltsov2015d}. GIO are operations
which preserve all incoherent states, i.e., $\Lambda[\varrho]=\varrho$
for any incoherent state $\varrho$. In particular, every GIO is incoherent
regardless of the particular Kraus decomposition, i.e., for every
experimental realization of the operation. Since GIO do not allow
for transformations between different incoherent states, notably for example between the energy eigenstates (when coherence is measured with respect to the eigenbasis of the Hamiltonian of the system),  they capture
the framework of coherence in the presence of additional constraints,
such as energy conservation. FIO is in turn the most general set of operations
which is incoherent for every Kraus decomposition \cite{Streltsov2015d}.
The GIO framework is closely related to the concept of \emph{resource
destroying maps} introduced by \cite{Liu2017}. The latter studies
quantum operations which transform any quantum state onto a free state,
and moreover preserve all free states. If the set of free states is
taken to be incoherent, any such coherence destroying map is also
GIO.

The role of energy in the context of coherence has also been investigated
by \cite{Yang2015}. In particular, \cite{Yang2015} defined the class
of \textit{energy preserving operations} (EPO) as all operations which
have a free dilation as in Eq.~(\ref{eq:Stinespring}), with the
additional requirement that the unitary commutes with the Hamiltonian
of the system and environment individually. Note that the set of EPO
is a strict subset of TIO \cite{Yang2015}.

That there is such a wide variety of possible definitions of incoherent
operations should not come as a surprise, as it mirrors the situation
in entanglement theory where the set LOCC is operationally well-defined,
though mathematically cumbersome, while larger sets such as separable
operations \cite{Vedral1998} and positive partial transpose preserving
operations \cite{Rains2001} have mathematically very convenient properties,
but in general do require a finite amount of free entanglement for
their implementation \cite{Bennett1999}. Nevertheless, they are very
useful as they provide bounds on achievable efficiencies for
transformations under the difficult to handle LOCC constraints. Equally,
the wealth of definitions for alternative incoherent operations can
be expected to provide both conceptual and quantitative insights into
the properties of coherence as a resource.

A classification of the different frameworks of coherence, motivated
by the notion of speakable and unspeakable information \cite{Peres2002,Bartlett2007},
has been proposed by \cite{Marvian2016}. Speakable information includes
any type of information for which the means of encoding is not relevant,
while unspeakable information depends on the way of encoding. In general,
a framework of coherence is speakable, if it allows for a free transformation
between the states $(\ket{0}+\ket{1})/\sqrt{2}$ and $(\ket{0}+\ket{2})/\sqrt{2}$.
Otherwise, the framework of coherence is unspeakable \cite{Marvian2016}.
In Table \ref{tab:Table} we summarize the aforementioned frameworks
of coherence according to this classification.

\begin{table}
\begin{tabular}{lllllll}
\hline
 & \multicolumn{6}{c}{Sets of free operations}\tabularnewline
\hline
Speakable coherence  & FIO  & PIO  & SIO  & IO  & DIO  & MIO\tabularnewline
Unspeakable coherence  & GIO  & EPO  & \multicolumn{1}{l}{TIO} & \multicolumn{1}{l}{} & \multicolumn{1}{l}{} & \tabularnewline
\hline
\end{tabular}

\caption{\label{tab:Table} Classification of different frameworks of coherence
into speakable and unspeakable type \cite{Marvian2016}.}
\end{table}

Before we move on, we would like to discuss two possible extensions
of incoherent operations which also apply to other classes that have
been presented above. First, motivated by observations in entanglement
theory \cite{Jonathan1999a} one might define the set of {\em catalytic
incoherent operations} \cite{Baumgratz2014} by allowing for the
use of an additional physical system of arbitrary dimension, the catalyst,
in a state $\eta$ which has to be returned unchanged after the transformation.
That is, for arbitrary states $\varrho$ and $\sigma$ of the system
and (bipartite) incoherent operations $\Lambda$ we require
\begin{equation}
\Lambda[\varrho\otimes\eta]=\sigma\otimes\eta.
\end{equation}
Indeed, in the theory of entanglement the addition of catalysts is
known to confer additional power in that it makes possible state transformations
that would otherwise be impossible under LOCC alone \cite{Jonathan1999a}.
As was shown by \cite{Bu2015b}, these results also hold in the resource theory of coherence: catalytic incoherent operations allow for pure state transformations which cannot be achieved with incoherent operations alone.
The role of catalysts for the class TIO has also been discussed
in \cite{Marvian2013,Marvian2016b}.

Concerning the second possible extension, up until this point we have
considered only exact state transformations, an idealization which,
arguably, cannot be achieved in the real world. Hence, already from
this practical consideration, one should also permit approximate state
transformations such that, instead of the exact map $\Lambda[\varrho]=\sigma$,
one would be satisfied with achieving $||\Lambda[\varrho]-\sigma||_{1}\le\epsilon$
for some small $\epsilon$, where $||M||_{1}=\mathrm{Tr}\sqrt{M^{\dagger}M}$
denotes the trace norm. Beyond the mere practical motivation, allowing
for approximate transformations has some relevance because it can
considerably change the set of available state transformations. In
particular, when permitting small errors to occur in catalysts their
power may increase further, in fact so much that any state transformation
may potentially become possible through embezzling of quantum states
\cite{vanDam2003}. Stronger constraints such as substituting $\epsilon$
by $\epsilon/\log D$ where $D=\text{dim}[{\eta}]$ and $\eta$ is
the state of the catalyst can prevent embezzling (see \cite{Brandao2015a}
for examples in quantum thermodynamics). On the other hand, the class
of {\em asymptotically exact incoherent operations} becomes particularly
relevant when one wishes to consider the mapping $\varrho\mapsto\sigma$
not for a single copy, but starting from $N$ copies, $\varrho^{\otimes N}$.
In this case, one typically asks for the minimal rate $r$ such that
$\lim_{N\rightarrow\infty}\{\inf_{\Lambda}||\Lambda[\varrho^{\otimes\left\lfloor rN\right\rfloor }]-\sigma^{\otimes N}||_{1}\}=0$,
where $r$ is the asymptotic cost of $\sigma$ in terms of $\varrho$,
and $\lfloor x\rfloor$ denotes the largest integer lower than or
equal to the real number $x$. Indeed, \cite{Winter2015} have carried
out such a program defining the appropriate quantities and obtaining
results that will be presented in more detail in Sec.~\ref{sub:Distillable-coherence}.
Approximate transformations for the class TIO have also been
investigated recently by \cite{Marvian2016b}.

\subsection{Coherence as a resource}

We now proceed to explore the use of coherence as a resource for enabling
operations that would not otherwise be possible if only incoherent
operations were accessible to the experimenter.

\subsubsection{\label{sec:maxcoh}Maximally coherent states and state transformations
via incoherent operations}

We start by identifying a $d$-dimensional maximally coherent state
as a state that allows for the \textit{deterministic} generation of
\textit{all} other $d$-dimensional quantum states by means of the free operations.
Note that this definition is independent of a specific
coherence quantifier and allows to identify a unit for coherence (coherence bit, or
{\it cobit})\footnote{The name ``cobit'' for a unit of coherence has been first used in \cite{Chitambar2015a}}
to which all measures may be normalized; see also Sec.~\ref{sec:Quantifying-coherence}
for more details on quantifying coherence. The canonical example of
a maximally coherent state is
\begin{equation}
|\Psi_{d}\rangle=\frac{1}{\sqrt{d}}\sum_{i=0}^{d-1}|i\rangle
\end{equation}
as it is easy to see that by IO, any $d$-dimensional state $\varrho$
may be prepared from $|\Psi_{d}\rangle$ with certainty \cite{Baumgratz2014}.
We note that not all frameworks of coherence that were discussed in
Sec.~\ref{sub:Classes} have a maximally coherent state, i.e., a
state from which all other states can be created via the corresponding
set of operations. In particular, such a state does not exist for
PIO \cite{Chitambar2016,Chitambar2016b,Chitambar2017}, GIO, and FIO \cite{Streltsov2015d}.

The full set of maximally coherent states is obtained as the orbit
of $|\Psi_{d}\rangle$ under all incoherent unitaries \cite{Bai2015}.
For coherence theory the maximally coherent state $|\Psi_{d}\rangle$
plays a role analogous to the maximally entangled state in entanglement
theory. As we shall see later, both concepts are very closely related.
One may also determine maximally coherent states under certain additional
constraints, such as the degree of mixedness, which gives rise to
the class of {\em maximally coherent mixed states} \cite{Singh2015}. For a $d$-dimensional system and any fixed spectrum $\{p_n\}_{n=1}^d$, \cite{Alex2016} have proven that the state $\varrho_{\max} = \sum_{n=1}^d  p_n\ket{n_+}\!\bra{n_+}$, where $\{\ket{n_+}\}$ denotes a mutually unbiased basis with respect to the incoherent basis $\{\ket{i}\}$ (that is, $\left|\langle i | n_+ \rangle \right|^2=\frac{1}{d}$), is a universal maximally coherent mixed state with respect to any coherence monotone under the set MIO.
Complementarity relations between measures of purity
and coherence have been further investigated in \cite{Xi2014,Cheng2015,Kumar2015,Giorda2016,Alex2016}.

The interconversion of pure states via incoherent operations has been
studied by \cite{Du2015b,Du2017,Winter2015,Chitambar2016,Chitambar2016b,Chitambar2017,Zhu2017}. In particular,
a pure state $\ket{\phi}$ can be converted into another pure state
$\ket{\psi}$ via IO or SIO if and only if $\Delta[\ket{\psi}\!\bra{\psi}]$ majorizes
$\Delta[\ket{\phi}\!\bra{\phi}]$ \cite{Winter2015,Zhu2017}, i.e.,
\begin{equation}
\Delta[\ket{\psi}\!\bra{\psi}]\succ\Delta[\ket{\phi}\!\bra{\phi}]
\end{equation}
is a necessary and sufficient condition for the transformation
$\ket{\phi}\mapsto\ket{\psi}$ via IO or SIO\footnote{For IO, this result was first claimed by \cite{Du2015b}, however the proof turned out to be incomplete \cite{Chitambar2016,Chitambar2016b,Chitambar2017,Winter2015}.
In a recent erratum, \cite{Du2017} addressed the criticism of
\cite{Chitambar2016,Chitambar2016b}, and presented an alternative proof
for systems of dimension~3. A complete proof for IO in any dimension was presented recently by \cite{Zhu2017}.}. Here, the majorization relation for
density matrices $\varrho\succ\sigma$ means that their spectra $\mathrm{spec}(\varrho)=(p_{1}\geq\ldots\geq p_{d})$
and $\mathrm{spec}(\sigma)=(q_{1}\geq\ldots\geq q_{d})$ fulfill the
relation $\sum_{i=1}^{t}p_{i}\geq\sum_{j=1}^{t}q_{j}$ for all $t<d$,
and $\sum_{i=1}^{d}p_{i}=\sum_{j=1}^{d}q_{j}$. Note that
a similar majorization condition is also known in entanglement theory
\cite{Nielsen1999}.

Less progress has been made for the single copy transformations of
mixed coherent states and only isolated results are known. A notable
result in this context was provided by \cite{Chitambar2016,Chitambar2016b,Chitambar2017}, who
gave a full characterization of single-qubit state conversion via
SIO, DIO, IO, or MIO. In particular, a single-qubit state with Bloch vector $\boldsymbol{r}=(r_x,r_y,r_z)^T$ can be converted into another single-qubit state with Bloch vector $\boldsymbol{s}=(s_x,s_y,s_z)^T$ via SIO, DIO, IO or MIO if and only if the following inequalities are fulfilled \cite{Streltsov2017}:
\begin{eqnarray}
s_x^2 + s_y^2 & \leq & r_x^2 + r_y^2, \\
\frac{1-s_z^2}{s_x^2 + s_y^2} & \geq & \frac{1-r_z^2}{r_x^2 + r_y^2}.
\end{eqnarray}

In the asymptotic setting, any state which is not incoherent allows
for the distillation of maximally coherent states via IO \cite{Winter2015}.
The optimal rate for this process can be evaluated analytically. We
refer to Sec.~\ref{sub:Distillable-coherence} for more details,
where also the asymptotic state formation from maximally coherent
states is discussed.

\subsubsection{States and maps}

A more complex task beyond state preparation is that of the generation
of a general quantum operation from a supply of coherent states and
incoherent operations. Just as the maximally entangled state allows
for the generation of all quantum operations \cite{Eisert2000} via
LOCC, so does the maximally coherent state allow for the generation
of all quantum operations via IO. The explicit construction for an
arbitrary single-qubit unitary can be found in \cite{Baumgratz2014},
and the extension to general quantum operations of arbitrary dimension was studied by \cite{Chitambar2015a,BenDana2017}. In particular, it was shown by \cite{BenDana2017} that any quantum operation acting on a Hilbert space of dimension $d$ can be implemented via IO by consuming a maximally coherent state of dimension $d$.
The corresponding construction makes use of maximally coherent states even if the targeted quantum operation may only be very slightly coherent. It is an open question how much coherence is required for creating
an arbitrary unitary, or an arbitrary quantum operation in general.
However, by virtue of the monotonicity of coherence quantifiers
under incoherent operations, lower bounds to the amount of coherence
required to implement a quantum operation can be provided \cite{Mani2015,Bu2015,BenDana2017}.

\subsection{\label{sub:LICC}Quantum coherence in distributed scenarios}

Based on the framework of LOCC known from entanglement theory \cite{Plenio2007,HorodeckiRMP09},
one can introduce the framework of \emph{local incoherent operations
and classical communication} (LICC) \cite{Streltsov2015c,Chitambar2015a}.
In both concepts one has two separated parties, Alice and Bob, who
are connected via a classical channel (such as a telephone). While
in the LOCC framework Alice and Bob are allowed to locally perform
any quantum operation which is compatible with the laws of quantum
mechanics, in the framework of LICC the parties are restricted to
local incoherent operations only\footnote{In the frameworks studied in \cite{Streltsov2015c,Chitambar2015a} local operations were restricted to the IO class, but other classes of local free operations can also be considered.}.

While LICC operations in general have a difficult mathematical structure,
it is possible to introduce the more general class of \emph{separable
incoherent} (SI) operations which has a simple mathematical form \cite{Streltsov2015c}:
\begin{equation}
\Lambda[\varrho^{AB}]=\sum_{i}(A_{i}\otimes B_{i})\varrho^{AB}(A_{i}^{\dagger}\otimes B_{i}^{\dagger}),\label{eq:SI}
\end{equation}
where $A_{i}$ and $B_{i}$ are local incoherent operators. General
quantum operations of the form~(\ref{eq:SI}), but without the incoherence
restriction, have been studied extensively in entanglement theory,
where they are called separable operations \cite{Vedral1998}. It
is an interesting open question whether the intersection of LOCC and
SI operations is equal to LICC operations \cite{Streltsov2015c}.

Asymmetric scenarios where only one of the parties is restricted to
IO locally have also been considered \cite{Chitambar2015,Streltsov2015c}.
In the case where the second party (Bob) is restricted to incoherent
operations, the corresponding sets of operations are called \emph{local
quantum-incoherent operations and classical communication} (LQICC)
and \emph{separable quantum-incoherent operations} (SQI) respectively.
LQICC operations are in particular important for the tasks of incoherent
quantum state merging \cite{Streltsov2016} (see also next Section)
and assisted coherence distillation \cite{Chitambar2015,Streltsov2015c}.
The latter task has also been performed experimentally \cite{Wu2017},
and will be discussed in more detail in Sec.~\ref{sub:Distillable-coherence-of-collaboration}.

A related set of operations has been presented by \cite{Matera2015}
in the context of a resource theory of control of quantum systems,
and called \emph{global operations incoherent on $B$} (GOI$B$).
Those operations allow for incoherent operations on the subsystem
$B$, and moreover they also allow for controlled operations in the
incoherent reference basis from $B$ to $A$. The latter amount to
arbitrary control unitaries of the form $U_{\mathrm{C}}=\sum_{i}U_{i}^{A}\otimes\ket{i}\!\bra{i}^{B}$
\cite{Matera2015}. Moreover, the framework also allows to attach
or discard subsystems of $A$, and to perform measurements with postselection
on this subsystem \cite{Matera2015}.

The sets LQICC, SQI, and GOI$B$ lead to the same set $\mathcal{QI}$
of free states, which are called \emph{quantum-incoherent states} \cite{Chitambar2015,Streltsov2015c,Matera2015} and take the form
\begin{equation}
\varrho=\sum_{i}p_{i}\sigma_{i}^{A}\otimes\ket{i}\!\bra{i}^{B}.\label{eq:qi}
\end{equation}
Here, $\sigma_{i}^{A}$ are arbitrary states on the subsystem $A$,
while $\ket{i}^{B}$ are incoherent states on the subsystem $B$.
Moreover, LQICC is a strict subset of SQI \cite{Streltsov2015c} and
GOI$B$ \cite{Matera2015}. In turn, the latter is a strict subset of
the maximal set of operations mapping the set of quantum-incoherent
states onto itself, defined in \cite{Ma2015}. We also note that GOI$B$
can create entanglement
whilst using up coherence \cite{Matera2015}, which is not possible
for LQICC and SQI operations. Finally, we note that the free states
under LICC and SI operations are bipartite incoherent states, i.e.,
convex combinations of incoherent product states \cite{Streltsov2015c}.

\subsection{Connection between coherence and entanglement theory}

The resource theory of coherence exhibits several connections to the
resource theory of entanglement. One of the first approaches in this
direction was presented in \cite{Streltsov2015b}, where it was shown
that any state with nonzero coherence can be used for entanglement
creation via bipartite incoherent operations; we refer to Sec.~\ref{sub:Coherence-Entanglement}
for a detailed discussion. These results were generalized to quantum
discord \cite{Ma2015} and general types of nonclassicality \cite{Killoran2015};
see Secs.~\ref{sub:Discord} and \ref{sub:General-measures-of-nonclassicality} for more details.

The interplay between coherence and entanglement in distributed scenarios
has been first studied by \cite{Chitambar2015a}, who investigated
state formation and distillation of entanglement and local coherence
via LICC operations. Interestingly, it was shown that a bipartite
state has distillable entanglement if and only if entanglement can
be distilled via LICC \cite{Chitambar2015a}.

Another relation between entanglement and coherence was provided in
\cite{Streltsov2016}, where the authors introduced and studied the
task of \emph{incoherent quantum state merging}. The latter is an
incoherent version of standard quantum state merging \cite{Horodecki2005a,Horodecki2007},
where two parties aim to merge their parts of a tripartite quantum
state while preserving correlations with a third party. While standard
quantum state merging can lead to a gain of entanglement \cite{Horodecki2005a,Horodecki2007},
no merging protocol can lead to a gain of entanglement and coherence
simultaneously \cite{Streltsov2016}.

Finally, we mention that the resource theory of coherence shares many
similarities with the resource theory of entanglement, if the latter
is restricted to maximally correlated states \cite{Winter2015,Chitambar2015}.
In particular, given a quantum state $\varrho=\sum_{ij}\varrho_{ij}\ket{i}\!\bra{j}$,
we can assign to it a bipartite maximally correlated state as $\varrho_{\mathrm{mc}}=\sum_{ij}\varrho_{ij}\ket{ii}\!\bra{jj}$.
An important open question in this context is whether any of the aforementioned
classes of incoherent operations $\Lambda_{\mathrm{i}}$ is equivalent
to LOCC operations $\Lambda_{\mathrm{LOCC}}$ on maximally correlated
states, i.e.,
\begin{equation}
\sigma=\Lambda_{\mathrm{i}}[\varrho]\overset{?}{\Leftrightarrow}\sigma_{\mathrm{mc}}=\Lambda_{\mathrm{LOCC}}[\varrho_{\mathrm{mc}}].
\end{equation}
Partial answers to this questions were presented for IO in \cite{Winter2015,Chitambar2015}
and for GIO in \cite{Streltsov2015d}.
As we will also discuss in the following Section, several quantifiers
of coherence known today coincide with their equivalent entanglement
quantifiers for the corresponding maximally correlated state.

\section{\label{sec:Quantifying-coherence}Quantifying quantum coherence}

\subsection{Postulates for coherence monotones and measures}

The first axiomatic approach to quantify coherence has been
presented by \cite{Aberg2006}, and an alternative framework has been
developed more recently by \cite{Baumgratz2014}. The latter approach can
be seen as parallel to the axiomatic quantification of entanglement,
first introduced two decades ago \cite{Vedral1997,Vedral1998}. The
basis of the axiomatic approach consists of the following postulates
that any quantifier of coherence $C$ should fulfill \cite{Aberg2006,Levi14,Baumgratz2014}:
\begin{lyxlist}{00.00.0000}
\item [{(C1)}] {\it Nonnegativity}:
\begin{equation}
C(\varrho)\geq0
\end{equation}
in general, with equality if and only if $\varrho$ is incoherent.
\item [{(C2)}] {\it Monotonicity}: $C$ does not increase under the action of
incoherent operations\footnote{While this condition was originally proposed for the set IO \cite{Baumgratz2014},
it can be generalized in a straightforward way to any set of operations
discussed in Sec.~\ref{sub:Classes}}, i.e.,
\begin{equation}
C(\Lambda[\varrho])\leq C(\varrho)
\end{equation}
for any incoherent operation $\Lambda$.
\item [{(C3)}] {\it Strong monotonicity}: $C$ does not increase on average under
selective incoherent operations, i.e.,
\begin{equation}
\sum_{i}q_{i}C(\sigma_{i})\leq C(\varrho)
\end{equation}
with probabilities $q_{i}=\mathrm{Tr}[K_{i}\varrho K_{i}^{\dagger}]$,
post-measurement states $\sigma_{i}=K_{i}\varrho K_{i}^{\dagger}/q_{i}$,
and incoherent Kraus operators $K_{i}$.
\item [{(C4)}] {\it Convexity}: $C$ is a convex function of the state, i.e.,
\begin{equation}
\sum_{i}p_{i}C(\varrho_{i})\geq C\left(\sum_{i}p_{i}\varrho_{i}\right).
\end{equation}

\end{lyxlist}
At this point it is instrumental to compare conditions C2 and C3 to
the corresponding conditions in entanglement theory \cite{Vedral1997,Vedral1998,Plenio2007,HorodeckiRMP09}.
There, C2 is equivalent to the requirement that an entanglement quantifier
$E$ should not increase under LOCC. Similarly, C3 is equivalent to
the requirement that $E$ should not increase on average under selective
LOCC operations, i.e., when the communicating parties are able to
store their measurement outcomes.

Conditions C1 and C2 can be seen as minimal requirements: any quantity
$C$ should at least fulfill these two conditions in order to be a
meaningful resource quantifier for some coherence-based task. Condition
C3 quantifies the intuition that coherence should not increase under
incoherent measurements even if one has access to the individual measurement
outcomes. This condition C3, when combined with convexity C4, implies
monotonicity C2 \cite{Baumgratz2014}. The reason for this overcompleteness
comes from entanglement theory: there are meaningful quantifiers of
entanglement which satisfy only some of the above conditions \cite{Plenio2007,HorodeckiRMP09}.

Following standard notions from entanglement theory \cite{Plenio2007},
we call a quantity $C$ which fulfills condition C1 and either condition
C2 or C3 (or both) a \emph{coherence monotone}. A quantity $C$ will
be further called a \emph{coherence measure} if it fulfills C1--C4
together with the following two additional conditions:
\begin{lyxlist}{00.00.0000}
\item [{(C5)}] {\it Uniqueness} for pure states: For any pure state $\ket{\psi}$
$C$ takes the form
\begin{equation}
C(\ket{\psi}\!\bra{\psi})=S(\Delta[\ket{\psi}\!\bra{\psi}]),\label{eq:C5}
\end{equation}
where $S(\varrho)=-\mathrm{Tr}[\varrho\log_{2}\varrho]$ is the von
Neumann entropy\footnote{Note that this condition may be considered too strong in view of the
fact that the quantity in the right-hand side of Eq.~(\ref{eq:C5}),
i.e.~the relative entropy of coherence (discussed in Sec.~\ref{sub:REC}),
adopts its operational meaning in terms of coherence distillation
and dilution only in the asymptotic limit. In a weaker form, postulate
C5 might read instead: $\forall\epsilon>0$ there exists a family
of states $\Phi(n)\in{\cal B}_{\epsilon}(\ket{\psi}\!\bra{\psi}^{\otimes n})$
such that $\lim_{n\rightarrow\infty}C(\Phi(n))/n=S(\Delta[\ket{\psi}\!\bra{\psi}])$.}.
\item [{(C6)}] {\it Additivity}: $C$ is additive under tensor products, i.e.,
\begin{equation}
C(\varrho\otimes\sigma)=C(\varrho)+C(\sigma).
\end{equation}

\end{lyxlist}
We wish to remark that the terminology adopted here differs
from the one used in some recent literature, which
is mainly based on \cite{Baumgratz2014}. In particular, several authors
require that a coherence measure only fulfills conditions C1--C4.
With the more stringent approach presented here, inspired from entanglement
theory, we aim to distinguish two important coherence quantifiers:
distillable coherence (which is equal to the relative entropy of coherence)
and coherence cost (which is equal to the coherence of formation).
As we will see in the following, these two quantities fulfill all
conditions C1--C6 and are thus elevated to the status of measures.
Most of the other coherence quantifiers presented in this Colloquium
remain coherence monotones, in the sense that they fulfill conditions
C1 and C2; several of them also turn out to fulfill C3 and C4, however
most violate C5 and C6.

Finally, we note that alternative or additional desirable requirements for
coherence monotones may be proposed. In particular, as was shown by \cite{Yu2016b},
for the set IO the conditions C3 and C4 can be replaced by the following
single requirement of additivity on block-diagonal states:
\begin{equation}\label{eq:cohaddoplus}
C\big(p\varrho\oplus(1-p)\sigma\big)=p C(\varrho)+(1-p)C(\sigma).
\end{equation}
When combining Eq.~(\ref{eq:cohaddoplus}) with C1 and C2, these three conditions are equivalent
to C1--C4 \cite{Yu2016b}.
Furthermore, \cite{Peng2016} postulated that any valid coherence quantifier should be maximal only on the set of pure maximally coherent states ${\ket{\Psi_{d}}}$,
a property satisfied in particular by the two coherence measures mentioned
just above; however, such states do not represent a golden
unit in all versions of the resource theory of coherence, as remarked
in Sec.~\ref{sec:maxcoh}.

In the following Sections, we review the most relevant coherence monotones and measures defined in current literature, as well as other recent quantitative studies of coherence --- and nonclassicality more broadly --- in single and multipartite quantum systems.

\subsection{\label{sub:Distillable-coherence}Distillable coherence and coherence
cost}

The {\em distillable coherence} is the optimal number of maximally
coherent single-qubit states $\ket{\Psi_{2}}$ which can be obtained
per copy of a given state $\varrho$ via incoherent operations in
the asymptotic limit. The formal definition of distillable coherence
can be given as follows \cite{Yuan2015,Winter2015}:
\begin{equation}
C_{d}(\varrho)=\sup\left\{ R:\lim_{n\rightarrow\infty}\left(\inf_{\Lambda_{\mathrm{i}}}\left\Vert \Lambda_{\mathrm{i}}[\varrho^{\otimes n}]-\ket{\Psi_{2}}\!\bra{\Psi_{2}}^{\otimes\left\lfloor nR\right\rfloor }\right\Vert _{1}\right)=0\right\} .\label{eq:Cd}
\end{equation}
From this definition it is tempting to believe that exact evaluation
of distillable coherence is out of reach. Surprisingly, this is not
the case, and a simple expression for distillable coherence of an
arbitrary mixed state, with $\Lambda_{\mathrm{i}}$ belonging to the
set IO, was given in \cite{Winter2015}:
\begin{equation}
C_{d}(\varrho)=S(\Delta[\varrho])-S(\varrho).\label{eq:Cd-1}
\end{equation}
For pure states this result was independently found by \cite{Yuan2015}.
Interestingly, the distillable coherence also coincides with the relative
entropy of coherence which was introduced in \cite{Baumgratz2014}
and will be discussed in Sec.~\ref{sub:REC}. Since the relative
entropy of coherence is a coherence measure \cite{Baumgratz2014},
the same is true for the distillable coherence, i.e., it fulfills
all requirements C1--C6.

Another important quantifier is the {\em coherence cost} \cite{Yuan2015,Winter2015}:
it quantifies the minimal rate of maximally coherent single-qubit
states $\ket{\Psi_{2}}$ required to produce a given state $\varrho$
via incoherent operations in the asymptotic limit. Its formal definition
is similar to that of entanglement cost, and can be given as follows:
\begin{equation}
C_{c}(\varrho)=\inf\left\{ R:\lim_{n\rightarrow\infty}\left(\inf_{\Lambda_{\mathrm{i}}}\left\Vert \varrho^{\otimes n}-\Lambda_{\mathrm{i}}\left[\ket{\Psi_{2}}\!\bra{\Psi_{2}}^{\otimes\left\lfloor nR\right\rfloor }\right]\right\Vert _{1}\right)=0\right\} .
\end{equation}
Interestingly, restricting again $\Lambda_{\mathrm{i}}$ to the set
IO, the coherence cost admits a single-letter expression. In particular,
it is equal to the coherence of formation \cite{Winter2015}:
\begin{equation}
C_{c}(\varrho)=C_{f}(\varrho).\label{eq:CcCf}
\end{equation}
The coherence of formation will be defined and discussed in detail
in Sec.~\ref{sub:Convex-roof-measures}. The result in Eq.~(\ref{eq:CcCf})
can be seen as parallel to the well known fact that the entanglement
cost is equal to the regularized entanglement of formation \cite{Hayden2001}.
However, in the case of coherence no regularization is required, which
significantly simplifies the evaluation of $C_{c}$.

The coherence cost is a coherence measure, i.e., it fulfills all conditions
C1--C6. Conditions C1--C4 follow from Eq.~(\ref{eq:CcCf}) and the
fact that the coherence of formation fulfills C1--C4 \cite{Yuan2015}.
Condition C5 also follows from Eq.~(\ref{eq:CcCf}) and the definition
of coherence of formation, see Section \ref{sub:Convex-roof-measures}.
Moreover, the coherence cost is additive, i.e., condition C6 is also
satisfied \cite{Winter2015}.

In general, the distillable coherence cannot be larger than the coherence
cost:
\begin{equation}
C_{d}(\varrho)\leq C_{c}(\varrho).
\end{equation}
For pure states this inequality becomes an equality, which implies
that the resource theory of coherence is reversible for pure states.
In particular, any state $\ket{\psi_{1}}$ with a distillable coherence
of $c_{1}$ cobits can be asymptotically converted into any other state
$\ket{\psi_{2}}$ with a distillable coherence of $c_{2}$ cobits at
a rate $c_{1}/c_{2}$. However, there exist mixed states with $C_{d}$
strictly smaller than $C_{c}$ \cite{Winter2015}. Nevertheless, this
phenomenon does not appear in its extremal form, i.e., there are no
states with zero distillable coherence but nonzero coherence cost
\cite{Winter2015}. Therefore, in contrast to entanglement theory
\cite{HorodeckiBound}, there is no `bound coherence' within the resource
theory of coherence based on the set IO \cite{Winter2015}.

Notice instead that, if one considers the maximal set MIO of incoherent
operations, the corresponding resource theory of coherence becomes
reversible also for mixed states, as the general framework of \cite{Brandao2015}
applies in this case \cite{Winter2015}. This means that the distillable
coherence under MIO remains equal to the relative entropy of coherence,
i.e.~Eq.~(\ref{eq:Cd-1}) still holds under the largest set of free
operations, while the coherence cost under MIO also reduces to the
same quantity, $C_{c}(\varrho)=C_{d}(\varrho)$, thus closing the
irreversibility gap\footnote{A similar situation occurs in the case of entanglement, for which
a fully reversible resource theory can also be constructed if one
considers the largest set of operations preserving separability \cite{Brandao2008,Brandao2010},
which is a strict superset of separable operations.}. We further note that not all sets of operations presented in Sec.~\ref{sub:Classes}
allow for free coherence distillation; this is further discussed in
Sec.~\ref{sec:conclusions}.

\subsection{\label{sub:Distance-based-measures}Distance-based quantifiers of
coherence}

A general distance-based coherence quantifier is defined as \cite{Baumgratz2014}
\begin{equation}
C_{D}(\varrho)=\inf_{\sigma\in{\cal I}}D(\varrho,\sigma),\label{eq:CDistance}
\end{equation}
where $D$ is a distance and the infimum is taken over the set of
incoherent states ${\cal I}$. Clearly, any quantity defined in this
way fulfills C1 for any distance $D(\varrho,\sigma)$ which is nonnegative
and zero if and only if $\varrho=\sigma$. Monotonicity C2 is also
fulfilled for any set of operations discussed in Sec.~\ref{sub:Classes}
if the distance is contractive \cite{Baumgratz2014},
\begin{equation}
D(\Lambda[\varrho],\Lambda[\sigma])\leq D(\varrho,\sigma)\label{eq:contractive}
\end{equation}
for any quantum operation $\Lambda$.

Moreover, any distance-based coherence quantifier fulfills convexity
C4 whenever the corresponding distance is jointly convex \cite{Baumgratz2014},
\begin{equation}
D\,\bigg(\sum_{i}p_{i}\varrho_{i},\sum_{j}p_{j}\sigma_{j}\bigg)\leq\sum_{i}p_{i}D\left(\varrho_{i},\sigma_{i}\right).\label{eq:joint-convexity}
\end{equation}
In the following, we will discuss explicitly three important distance-based
coherence quantifiers.

\subsubsection{\label{sub:REC} Relative entropy of coherence}

If the distance is chosen to be the quantum relative entropy
\begin{equation}
S(\varrho||\sigma)=\mathrm{Tr}[\varrho\log_{2}\varrho]-\mathrm{Tr}[\varrho\log_{2}\sigma],
\end{equation}
the corresponding quantifier is known as the {\em relative entropy
of coherence}\footnote{The relative entropy of coherence defined in Eq.~(\ref{eq:Cr})
coincides with a special instance of the {\it relative entropy of superposition}
\cite{Aberg2006} and of the {\it relative entropy of asymmetry} \cite{Vaccaro2008,Gour2009},
which will be discussed in Sec.~\ref{sub:Nonclassicality}. Furthermore, the quantity
in the right-hand side of Eq.~(\ref{eq:CrCd}) was independently proposed as a coherence
quantifier  by \cite{Herbut2005} under the name {\it coherence information}.} \cite{Baumgratz2014}:
\begin{equation}\label{eq:Cr}
C_{r}(\varrho)=\min_{\sigma\in{\cal I}}S(\varrho||\sigma).
\end{equation}
The relative entropy of coherence fulfills conditions C1, C4, and
C2 for any set of operations discussed in Sec.~\ref{sub:Classes}.
For the set IO it also fulfills C3 \cite{Baumgratz2014}. Moreover,
it is equal to the distillable coherence $C_{d}$, therefore both quantities admit the same closed expression \cite{Baumgratz2014,Winter2015,Gour2009}
\begin{equation}
    C_{r}(\varrho)=C_{d}(\varrho)=S(\Delta[\varrho])-S(\varrho),\label{eq:CrCd}
\end{equation}
where $\Delta[\varrho]$ is the dephasing operation defined in Eq.~(\ref{eq:dephasing}).
The proof of this equality is remarkably simple: it is enough to note
that the relative entropy between an arbitrary state $\varrho$ and
another incoherent state $\sigma\in{\cal I}$ can be written as
\begin{equation}
S(\varrho||\sigma)=S(\Delta[\varrho])-S(\varrho)+S(\Delta[\varrho]||\sigma),\label{eq:proof}
\end{equation}
which is straightforward to prove using the relation $\mathrm{Tr}[\varrho\log_{2}\sigma]=\mathrm{Tr}[\Delta[\varrho]\log_{2}\sigma]$.
Minimizing the right-hand side of Eq.~(\ref{eq:proof}) over all
incoherent states $\sigma\in{\cal I}$ we immediately see that the
minimum is achieved for $\sigma=\Delta[\varrho]$, which completes
the proof of Eq.~(\ref{eq:CrCd}). From Eq.~(\ref{eq:CrCd}) it
also follows that the relative entropy of coherence fulfills conditions
C5 and C6.

As was further shown by \cite{Singh2015b}, the relative entropy of
coherence can also be interpreted as the minimal amount of noise required
for fully decohering the state $\varrho$. Moreover, it was shown
in \cite{Rodrigues-Rosario2013} that the relative entropy of coherence
is related to the deviation of the state from thermal equilibrium.

Possible extensions of the relative entropy of coherence to quantifiers based
on the relative R\'{e}nyi and Tsallis entropies have also been discussed \cite{Chitambar2016,Chitambar2016b,Chitambar2017,Rastegin2016}.

\subsubsection{Coherence quantifiers based on matrix norms}

\label{sub:l1}

We will now consider coherence quantifiers based on matrix norms,
i.e., such that the corresponding distance has the form $D(\varrho,\sigma)=||\varrho-\sigma||$
with some matrix norm $||\cdot||$. We first note that any such distance
is jointly convex, i.e., fulfills Eq.~(\ref{eq:joint-convexity}),
as long as the corresponding norm $||\cdot||$ fulfills the triangle
inequality and absolute homogeneity\footnote{Absolute homogeneity of a matrix norm $||\cdot||$ means that $||aM||=|a|\times||M||$
holds for any complex number $a$ and any matrix $M$.} \cite{Baumgratz2014}. Thus, any matrix norm with the aforementioned
properties gives rise to a convex coherence quantifier. Relevant norms
with these properties are the $l_{p}$-norm $||\cdot||_{l_{p}}$ and
the Schatten $p$-norm $||\cdot||_{p}$:
\begin{align}
\left\Vert M\right\Vert _{l_{p}} & =\left(\sum_{i,j}\left|M_{ij}\right|^{p}\right)^{1/p},\\
\left\Vert M\right\Vert _{p} & =\left(\mathrm{Tr}\left[\left(M^{\dagger}M\right)^{p/2}\right]\right)^{1/p}
\end{align}
with $p\geq1$. The corresponding coherence quantifiers will be denoted
by $C_{l_{p}}$ and $C_{p}$ respectively.

The {\em $l_{1}$-norm of coherence} $C_{l_{1}}$ was first introduced
and studied in \cite{Baumgratz2014}. In particular, $C_{l_{1}}$
fulfills the conditions C1--C4 for the set IO and has the following
simple expression \cite{Baumgratz2014}:
\begin{equation}
C_{l_{1}}(\varrho)=\min_{\sigma\in{\cal I}}\left\Vert \varrho-\sigma\right\Vert _{l_{1}}=\sum_{i\neq j}|\varrho_{ij}|.
\end{equation}
For the maximally coherent state $C_{l_{1}}$ takes the value $C_{l_{1}}(\ket{\Psi_{d}})=d-1$,
which also means that $C_{l_{1}}$ does not fulfill conditions C5
and C6. It has also been shown in \cite{Bu2016b} that $C_{l_{1}}$ is a DIO and MIO monotone for $d=2$, while it violates the monotonicity condition C2 for DIO and MIO for $d>2$.

For $p=1$ the corresponding Schatten norm reduces to the trace norm.
Since the trace norm is contractive under quantum operations, the
corresponding coherence quantifier $C_{1}$ satisfies the conditions
C1, C4, and C2 for any set of operations discussed in Sec.~\ref{sub:Classes}.
However, $C_{1}$ violates the condition C3 for the set IO,
as was shown in \cite{Yu2016b} based on the violation of property in
Eq.~(\ref{eq:cohaddoplus}). Nevertheless,  for single-qubit states
$C_{1}$ is equivalent to $C_{l_{1}}$, and thus the condition C3
is fulfilled in this case \cite{Shao2015}. Similar results were
obtained for general $X$-states: also in this case $C_{1}$ is equivalent
to $C_{l_{1}}$, and condition C3 is fulfilled for the set IO \cite{Rana2015}. The characterization of the closest incoherent state with respect to the trace norm is nontrivial in general, and partial results for pure states and $X$-states have been obtained in \cite{Rana2015,Chen2016}.

For $p=2$ the Schatten norm is equivalent to the Hilbert-Schmidt
norm, and also equal to the $l_{2}$-norm. In this context, \cite{Baumgratz2014}
studied the case where coherence is quantified via the squared Hilbert-Schmidt
norm. They found that the coherence quantifier obtained in this way
violates strong monotonicity C3 for the set IO. For any $p\geq1$,
$C_{p}$ is a convex coherence monotone for all single-qubit states,
i.e., it fulfills C1, C4, and C2 for the set IO \cite{Rana2015}.
However, in general $C_{l_{p}}$ and $C_{p}$ violate conditions C2
and C3 for the set IO for all $p>1$ in higher-dimensional systems
\cite{Rana2015}. Interestingly, $C_{p}$ is a coherence monotone
for the set GIO for all $p\geq1$ \cite{Streltsov2015d}.

\subsubsection{\label{sub:GMC}Geometric coherence}

Here we will consider the {\em geometric coherence} defined in
\cite{Streltsov2015b} as follows:
\begin{equation}
C_{g}(\varrho)=1-\max_{\sigma\in\mathcal{I}}F(\varrho,\sigma)\label{eq:CMG}
\end{equation}
with the fidelity $F(\varrho,\sigma)=||\sqrt{\varrho}\sqrt{\sigma}||_{1}^{2}$.
The geometric coherence fulfills conditions C1, C4, and C2 for any
set of operations discussed in Sec.~\ref{sub:Classes}. Additionally,
it also fulfills C3 for the set IO \cite{Streltsov2015b}. For pure
states, the geometric coherence takes the form $C_{g}(\ket{\psi})=1-\max_{i}|\!\braket{i|\psi}\!|^{2}$,
which means that $C_{g}$ does not meet conditions C5 and C6.

If $\varrho$ is a single-qubit state, $C_{g}$ admits the following
closed expression \cite{Streltsov2015b}:
\begin{equation}
C_{g}(\varrho)=\frac{1}{2}\left(1-\sqrt{1-4|\varrho_{01}|^{2}}\right),\label{eq:CMG-qubit}
\end{equation}
where $\varrho_{01}=\braket{0|\varrho|1}$ is the offdiagonal element
of $\varrho$ in the incoherent basis. Note that for all single-qubit
states we have $C_{1}=C_{l_{1}}=2|\varrho_{01}|$, and thus $C_{g}$
is a simple function of these quantities in the single-qubit case.
As we will see in the following subsections, $C_{g}$ can also be
considered as a convex roof quantifier of coherence, and is also closely
related to the geometric entanglement. Upper and lower bounds on
the geometric coherence have been investigated in \cite{Zhang2017}.

Another related quantity was introduced by \cite{Baumgratz2014} and
studied in \cite{Shao2015} where it was called {\em fidelity of
coherence}: $C_{F}(\varrho)=1-\max_{\sigma\in{\cal I}}\sqrt{F(\varrho,\sigma)}$.
While $C_{F}$ fulfills conditions C1, C4, and C2 for any set of operations
discussed in Sec.~\ref{sub:Classes}, it violates the strong monotonicity
condition C3 for IO \cite{Shao2015}.

\subsection{Convex roof quantifiers of coherence\label{sub:Convex-roof-measures}}

Provided that a quantifier of coherence has been defined for all pure
states, it can be extended to mixed states via the standard convex
roof construction \cite{Yuan2015}:
\begin{equation}
C(\varrho)=\inf_{\{p_{i},\ket{\psi_{i}}\}}\sum_{i}p_{i}C(\ket{\psi_{i}}),
\end{equation}
where the infimum is taken over all pure state decompositions of $\varrho=\sum_{i}p_{i}\ket{\psi_{i}}\!\bra{\psi_{i}}$.
Constructions of this type have been previously introduced and widely
studied in entanglement theory \cite{Bennett1996,Uhlmann1998}.

If for pure states the amount of coherence is quantified by the distillable
coherence $C_{d}(\ket{\psi})=S(\Delta[\ket{\psi}\!\bra{\psi}])$,
as in Eq.~(\ref{eq:C5}), the corresponding convex roof quantifier
is known as the \emph{coherence of formation} \cite{Aberg2006,Yuan2015,Winter2015}:
\begin{equation}
C_{f}(\varrho)=\inf_{\{p_{i},\ket{\psi_{i}}\}}\sum_{i}p_{i}S(\Delta[\ket{\psi_{i}}\!\bra{\psi_{i}}]).\label{eq:Cf}
\end{equation}
As already noted in Sec.~\ref{sub:Distillable-coherence}, $C_{f}$
is equal to the coherence cost under IO. As was shown by \cite{Yuan2015},
the coherence of formation\footnote{\cite{Yuan2015} called this quantity intrinsic randomness.}
fulfills conditions C1--C4 for the set
IO. By definition, $C_{f}$ also fulfills C5, and additivity C6 was
proven in \cite{Winter2015}. Remarkably, $C_f$ violates monotonicity C2 for the class MIO \cite{Hu2016b}.

For a general state $\varrho=\sum_{ij}\varrho_{ij}\ket{i}\!\bra{j}$,
the coherence of formation is equal to the entanglement of formation
\cite{Bennett1996} of the corresponding maximally correlated state
$\varrho_{\mathrm{mc}}=\sum_{ij}\varrho_{ij}\ket{ii}\!\bra{jj}$ \cite{Winter2015}:
\begin{equation}
C_{f}(\varrho)=E_{f}(\varrho_{\mathrm{mc}}).\label{eq:CfEf}
\end{equation}
Using the formula for the entanglement of formation of two-qubit states
\cite{Wootters1998}, Eq.~(\ref{eq:CfEf}) implies that the coherence
of formation can be evaluated exactly for all single-qubit states~\cite{Yuan2015}:
\begin{equation}\label{eq:CfEf2q}
C_{f}(\varrho)=h\left(\frac{1+\sqrt{1-4|\varrho_{01}|^{2}}}{2}\right),
\end{equation}
where $h(x)=-x\log_{2}x-(1-x)\log_{2}(1-x)$ is the binary entropy.
If we compare this expression with the expression for the geometric
coherence in Eq.~(\ref{eq:CMG-qubit}), it follows that $C_{f}=h(1-C_{g})$
holds for any single-qubit state. Thus, $C_{f}$ is also a simple
function of $C_{1}$ and $C_{l_{1}}$ in this case.

The {\em coherence concurrence} was defined in \cite{CoherenceConcurrence} as the convex roof of the $l_1$-norm of coherence,
\begin{equation}
C_{c_1}(\varrho)=\inf_{\{p_{i},\ket{\psi_{i}}\}}\sum_{i}p_{i}C_{l_1}(\ket{\psi_{i}}\!\bra{\psi_{i}}).\label{eq:Cc1}
\end{equation}
It follows by definition that $C_{c_1}(\ket{\psi}\!\bra{\psi})=C_{l_1}(\ket{\psi}\!\bra{\psi})$ for any pure state $\ket{\psi}$, while $C_{c_1}(\varrho) \geq C_{l_1}(\varrho)$ for an arbitrary mixed state $\varrho$.
The coherence concurrence satisfies properties C1--C4 for the set IO as proven in \cite{CoherenceConcurrence}, but it violates C5 and C6. The relation between coherence concurrence and entanglement concurrence \cite{Wootters1998} was also investigated in \cite{CoherenceConcurrence}. In particular, for a single-qubit state $\varrho$, $C_{c_1}(\varrho)=C_{l_1}(\varrho)=2|\varrho_{01}|$, which means that the coherence concurrence of $\varrho$ is equal to the entanglement concurrence of the corresponding maximally correlated two-qubit state $\varrho_{\mathrm{mc}}$, and that the functional relation between $C_f$ and $C_{c_1}$ for a qubit, obtained from Eq.~(\ref{eq:CfEf2q}), is the same as that between the entanglement of formation and the entanglement concurrence for two qubits, established in \cite{Wootters1998}.

The geometric coherence $C_{g}$ introduced in Eq.~(\ref{eq:CMG})
can also be regarded as a convex roof quantifier:
\begin{equation}
C_{g}(\varrho)=\inf_{\{p_{i},\ket{\psi_{i}}\}}\sum_{i}p_{i}C_{g}(\ket{\psi_{i}}).
\end{equation}
This can be proven using a general theorem in \cite{Streltsov2010}
(see Theorem 2 in the appendix there). Moreover, Eq.~(\ref{eq:CfEf})
also holds for the geometric coherence and entanglement: $C_{g}(\varrho)=E_{g}(\varrho_{\mathrm{mc}})$,
where the geometric entanglement is defined as $E_{g}(\varrho)=1-\max_{\sigma\in{\cal S}}F(\varrho,\sigma)$,
and ${\cal S}$ is the set of separable states \cite{Wei2003,Streltsov2010}.

\subsection{\label{sub:Coherence-Entanglement}Coherence monotones from entanglement}

An alternative approach to quantify coherence has been introduced
in \cite{Streltsov2015b}. In particular, the authors showed that
any entanglement monotone $E$ gives rise to a coherence monotone
$C_{E}$ via the following relation:
\begin{equation}
C_{E}(\varrho^{S})=\lim_{d_{A}\rightarrow\infty}\left\{ \sup_{\Lambda_{\mathrm{i}}}E^{S:A}\left(\Lambda_{\mathrm{i}}\left[\varrho^{S}\otimes\ket{0}\bra{0}^{A}\right]\right)\right\} .\label{eq:CE}
\end{equation}
Here, $\varrho^{S}$ is a state of the system $S$, and $A$ is an
additional ancilla of dimension $d_{A}$. The supremum is performed
over all bipartite incoherent operations $\Lambda_{\mathrm{i}}\in$
IO, i.e., such that the corresponding Kraus operators map the product
basis $\ket{k}\ket{l}$ onto itself.

The intuition behind the {\em entanglement-based coherence quantifiers}
$C_{E}$ is the following: if the state $\varrho^{S}$ is incoherent,
then the total state $\Lambda_{\mathrm{i}}[\varrho^{S}\otimes\ket{0}\bra{0}^{A}]$
will remain separable for any bipartite incoherent operation $\Lambda_{\mathrm{i}}$.
However, if the state $\varrho^{S}$ has nonzero coherence, some incoherent
operation $\Lambda_{\mathrm{i}}$ can in fact create entanglement
between the system $S$ and the ancilla $A$ (see also Fig.~\ref{fig:plan},
right panel). This finding is quantified in Eq.~(\ref{eq:CE}): $C_{E}$
is a coherence monotone whenever $E$ is an entanglement monotone.
In particular, $C_{E}$ fulfills conditions C1--C4 for the set IO
whenever $E$ fulfills the corresponding conditions in entanglement
theory \cite{Streltsov2015b}.

In various situations $C_{E}$ admits an explicit formula. In particular,
if $E$ is the distillable entanglement, $C_{E}$ amounts to the distillable
coherence \cite{Streltsov2015b}:
\begin{equation}
C_{E_{d}}(\varrho)=C_{d}(\varrho).
\end{equation}
If $E$ is the relative entropy of entanglement, $C_{E}$ is the relative
entropy of coherence, which again is equal to $C_{d}$. A similar
relation can be found for the geometric entanglement and the geometric
coherence: $C_{E_{g}}(\varrho)=C_{g}(\varrho)$ \cite{Streltsov2015b}.
For distillable entanglement, relative entropy of entanglement, and
geometric entanglement, the supremum in Eq.~(\ref{eq:CE}) is achieved
when $\Lambda_{\mathrm{i}}$ is the generalized CNOT operation, i.e.,
the optimal incoherent operation is the unitary
\begin{equation}
U_{\mathrm{CNOT}}\ket{i}\ket{0}=\ket{i}\ket{i}.\label{eq:CNOT}
\end{equation}
It is not known if this unitary is the optimal incoherent operation
for all entanglement monotones $E$.

If entanglement is quantified via the distance-based approach \cite{Vedral1997},
\begin{equation}
E_{D}(\varrho)=\inf_{\sigma\in{\cal S}}D(\varrho,\sigma)
\end{equation}
with a contractive distance $D$, it was further shown in \cite{Streltsov2015b}
that the creation of entanglement from a state $\varrho^{S}$ via
incoherent operations is bounded above by its distance-based coherence:
\begin{equation}
E_{D}^{S:A}\left(\Lambda_{\mathrm{i}}\left[\varrho^{S}\otimes\ket{0}\bra{0}^{A}\right]\right)\leq C_{D}\left(\varrho^{S}\right).
\end{equation}
While this result was originally proven for $\Lambda_{\mathrm{i}}\in\mathrm{IO}$
in \cite{Streltsov2015b}, it generalizes to any set of incoherent
operations presented in Sec.~\ref{sub:Classes}. In case the distance
$D$ is chosen to be the quantum relative entropy, there exists an
incoherent operation saturating the bound for any state $\varrho^{S}$
as long as $d_{A}\geq d_{S}$. The same is true if the distance is
chosen as $D(\varrho,\sigma)=1-F(\varrho,\sigma)$. In both cases,
the bound is saturated by the generalized CNOT operation, see Eq.~(\ref{eq:CNOT}).
These results also imply that a state $\varrho^{S}$ can be used to
create entanglement via incoherent operations if and only if $\varrho^{S}$
has nonzero coherence \cite{Streltsov2015b}.

\subsection{\label{sub:Robustness}Robustness of coherence}

Another coherence monotone was introduced by \cite{Napoli2016,Piani2016},
and termed {\em robustness of coherence}. For a given state $\varrho$,
it quantifies the minimal mixing required to make the state incoherent:
\begin{equation}
R_{C}(\varrho)=\min_{\tau}\left\{ s\geq0\left|\frac{\varrho+s\tau}{1+s}\in\mathcal{I}\right.\right\} ,\label{eq:Robustness}
\end{equation}
where the minimum is taken over all quantum states $\tau$. A similar
quantity was studied earlier in entanglement theory under the name
robustness of entanglement \cite{Vidal1999a,Steiner2003}. The robustness
of coherence fulfills C1, C4, and C2 for all sets of operations discussed
in Sec.~\ref{sub:Classes}, and additionally it also fulfills C3
for IO \cite{Napoli2016,Piani2016}. Moreover, $R_{C}$ coincides
with the $l_{1}$-norm of coherence for all single-qubit states, all
$X$-states, and all pure states. The latter result also implies that
$R_{C}$ does not comply with C5 and C6.

The robustness of coherence has an operational interpretation which
is related to the notion of {\em coherence witnesses}. A coherence
witness is defined in a similar way as an entanglement witness in
entanglement theory: it is a Hermitian operator $W$ such that $\mathrm{Tr}[W\sigma]\geq0$
is true for all incoherent states $\sigma\in{\cal I}$. Under the
additional constraint $W\leq\openone$ it was shown in \cite{Napoli2016,Piani2016}
that the following inequality holds true:
\begin{equation}
R_{C}(\varrho)\geq\max\left\{ 0,-\mathrm{Tr}[\varrho W]\right\} .
\end{equation}
Interestingly, for any state $\varrho$ there exists a witness $W$
saturating this inequality. On the one hand, this result means that
the robustness of coherence is accessible in laboratory by measuring
the expectation value of a suitable witness $W$, as recently demonstrated in a photonic experiment \cite{Wang2017}.
On the other hand,
it also means that $R_{C}$ can be evaluated via a semidefinite program.
The robustness of coherence is moreover a figure of merit in the task
of quantum phase discrimination; we refer to Sec.~\ref{sec:discr}
for its definition and detailed discussion. Moreover, the results
presented in \cite{Napoli2016,Piani2016} also carry over to the resource
theory of asymmetry, which is discussed in Sec.~\ref{sub:Asymmetry}.

\subsection{Coherence quantifiers from interferometric visibility}
Clearly, quantum coherence is required for the observation of interference
patterns, e.g.~in the double slit experiment. Recently, this idea
was formalized in \cite{Prillwitz2014}, where the authors studied
the problem to determine coherence properties from interference patterns.
Similar ideas were also put forth in \cite{Bera2015}, where the authors
studied the role of the $l_{1}$-norm of coherence in general multislit
experiments.
The authors of \cite{Bagan2016} derived two exact complementarity
relations between quantifiers of coherence and path information in
a multipath interferometer, using respectively the $l_{1}$-norm and
the relative entropy of coherence. Studies on the quantification of
interference and its relationship with coherence have also been performed
earlier, see e.g.~\cite{Braun2006}.

The authors of \cite{Biswas2017} recently presented a general framework to quantify coherence from visibility in interference phenomena.
Consider a multi-path interferometer, in which a single particle can be in one of $d$ paths, denoting the
path variable by a set of orthogonal vectors $\{\ket{j}\}_{j=0}^{d-1}$ which define the reference basis. If a local phase shift $\varphi_j$ is applied in each arm, the output state of the particle can be written as $\varrho(\vec{\varphi}) = U(\vec{\varphi})\varrho U^\dagger(\vec{\varphi})$, where $\varrho$ is the input state and $U(\vec{\varphi})=\sum_j e^{i\varphi_j} \ket{j}\!\bra{j}$. Then, by placing an output detector which implements a  measurement described by a positive-operator valued measure $M=(M_\omega)$, one observes outcomes $\omega$ sampled from the Born probability $p_{M|\varrho}(\omega|\vec{\varphi}) = \mathrm{Tr}[\varrho(\vec{\varphi}) M_\omega]$, which constitute the interference pattern.
One can then define suitable {\it visibility} functionals $V[p_{M|\varrho}]$, which intuitively capture the degree of variability of the interference pattern as a function of the phases $\{\varphi_j\}$. It was then shown in \cite{Biswas2017} that, for any visibility functional $V$ satisfying certain physical requirements, the corresponding optimal visibility (maximized over all output measurements $M$), defines a valid {\it interference-based coherence quantifier},
\begin{equation}\label{eq:CV}
C_V(\varrho) = \sup_M V[p_{M|\varrho}],
\end{equation}
which satisfies properties C1--C3 for the set SIO, and also C4 if $V$ is convex in $p$. Various examples of visibility functionals and the corresponding coherence monotones were further discussed in \cite{Biswas2017}, showing in particular that the robustness of coherence presented in Sec.~\ref{sub:Robustness} can be alternatively interpreted (up to a normalization factor) as an interference-based quantifier of the form (\ref{eq:CV}) for a suitable $V$, and that one can also obtain variants of the asymmetry monotones such as the Wigner-Yanase skew information discussed in Sec.~\ref{sub:Asymmetry}, which satisfy the necessary monotonicity requirements for coherence (with respect to the set SIO). These results establish important links between the somehow more abstract aspects in the resource theory of coherence and operational
notions in the physics of interferometers.

\bigskip

\subsection{Coherence of assistance}

The {\em coherence of assistance} was introduced by \cite{Chitambar2015}
as follows:
\begin{equation}
C_{a}(\varrho)=\sup_{\{p_{i},\ket{\psi_{i}}\}}\sum_{i}p_{i}S(\Delta[\ket{\psi_{i}}\!\bra{\psi_{i}}]),
\end{equation}
where the maximum is taken over all pure state decompositions of $\varrho=\sum_{i}p_{i}\ket{\psi_{i}}\!\bra{\psi_{i}}$.
This quantity is dual to the coherence of formation defined in Eq.~(\ref{eq:Cf})
as the minimum over all decompositions. If the state $\varrho$ is
the maximally mixed single-qubit state, it can be written as a mixture
of the maximally coherent states $\ket{\pm}=(\ket{0}\pm\ket{1})/\sqrt{2}$
with equal probabilities. For this reason we get $C_{a}(\openone/2)=1$,
i.e., the coherence of assistance is maximal for the maximally mixed
state. On the one hand, this means that the coherence of assistance
is not a coherence monotone, as it indeed violates condition C1. On
the other hand, the coherence of assistance plays an important role
for the task of assisted coherence distillation, see Sec.~\ref{sub:Distillable-coherence-of-collaboration}
for a detailed discussion.

While finding a closed expression for the coherence of assistance
seems difficult in general, its regularization admits a simple expression
\cite{Chitambar2015}:
\begin{equation}
C_{a}^{\infty}(\varrho)=\lim_{n\rightarrow\infty}\frac{1}{n}C_{a}\left(\varrho^{\otimes n}\right)=S(\Delta[\varrho]).
\end{equation}
Moreover, for all single-qubit states $\varrho$ the coherence of
assistance is $n$-copy additive, and thus can be written as $C_{a}(\varrho)=S(\Delta[\varrho])$.
There is further a close relation between the coherence of assistance
and the entanglement of assistance introduced in \cite{DiVincenzo1999}.
In particular, the coherence of assistance of a state $\varrho=\sum_{ij}\varrho_{ij}\ket{i}\!\bra{j}$
is equal to the entanglement of assistance of the corresponding maximally
correlated state $\varrho_{\mathrm{mc}}=\sum_{ij}\varrho_{ij}\ket{ii}\!\bra{jj}$
\cite{Chitambar2015}:
\begin{equation}
C_{a}(\varrho)=E_{a}(\varrho_{\mathrm{mc}}).
\end{equation}

\subsection{Coherence and quantum correlations beyond entanglement}
\label{sub:Discord}

Quantum discord \cite{Zurek2000,Ollivier2001,Henderson2001} is a
measure of quantum correlations going beyond entanglement \cite{Oppenheim2002,Modi2012,Streltsov2014,ABC}.
A bipartite quantum state $\varrho^{AB}$ is said to have zero discord
(with respect to the subsystem $A$) if and only if it can be written
as
\begin{equation}
\varrho_{\mathrm{cq}}^{AB}=\sum_{i}p_{i}\ket{e_{i}}\bra{e_{i}}^{A}\otimes\varrho_{i}^{B},\label{eq:cq}
\end{equation}
where the states $\{\ket{e_{i}}\}$ form an orthonormal basis, but
are not necessarily incoherent. The states of Eq.~(\ref{eq:cq})
are also known as classical-quantum \cite{Piani2008}, and the corresponding
set will be denoted by $\mathcal{CQ}$. Correspondingly, a state $\varrho^{AB}$
is called (fully) classically correlated, or classical-classical,
if and only if it can be written as \cite{Piani2008}
\begin{equation}
\varrho_{\mathrm{cc}}^{AB}=\sum_{i,j}p_{ij}\ket{e_{i}}\bra{e_{i}}^{A}\otimes\ket{e_{j}}\bra{e_{j}}^{B},
\end{equation}
and the corresponding set will be denoted by $\mathcal{CC}$. If a
state is not fully classically correlated, we say that it possesses
nonzero general quantum correlations. This notion can be straightforwardly
extended to more than two parties \cite{Piani2011}.

The amount of discord in a given state can be quantified via a distance-based
approach similar to the distance-based approach for coherence presented
in Sec.~\ref{sub:Distance-based-measures}. In particular, one can
define general distance-based quantifiers of discord $\delta_{D}^{A|B}$
and quantumness $Q_{D}^{A|B}$ as follows \cite{ABC,roga2015geometric,Modi2012}:
\begin{align}
\delta_{D}^{A|B}(\varrho^{AB}) & =\inf_{\sigma^{AB}\in\mathcal{CQ}}D(\varrho^{AB},\sigma^{AB}),\\
Q_{D}^{A|B}(\varrho^{AB}) & =\inf_{\sigma^{AB}\in\mathrm{\mathcal{CC}}}D(\varrho^{AB},\sigma^{AB})
\end{align}
with some distance $D$. If $D$ is chosen to be the quantum relative
entropy, the corresponding quantities are respectively known as the
relative entropy of discord and of quantumness \cite{Modi2010,Piani2011}.

Recently, \cite{Ma2015} studied the role of coherence for creating
general quantum correlations. The authors showed that the creation
of general quantum correlations from a state $\varrho$ is bounded
above by its coherence. In particular, if $D$ is a contractive distance,
the following relation holds for any pair of quantifiers $Q_{D}$
and $C_{D}$ \cite{Ma2015}:
\begin{equation}
Q_{D}^{S|A}\left(\Lambda_{\mathrm{i}}\left[\varrho^{S}\otimes\sigma^{A}\right]\right)\leq C_{D}\left(\varrho^{S}\right).\label{eq:discord}
\end{equation}
Here, $S$ is a system in an arbitrary state $\varrho^{S}$ and $A$
is an ancilla in an incoherent state $\sigma^{A}$. While this result
was originally proven for $\Lambda_{\mathrm{i}}\in$ IO in \cite{Ma2015},
it generalizes to any set of operations discussed in Sec.~\ref{sub:Classes}.
These results are parallel to the discussion on creation of entanglement
from coherence presented in \cite{Streltsov2015b}, see also Sec.~\ref{sub:Coherence-Entanglement}.
In particular, any state $\varrho^{S}$ with nonzero coherence can
be used for creating discord via incoherent operations, since any
such state can be used for the creation of entanglement \cite{Streltsov2015b}.

A general framework to define quantifiers of discord and quantumness
in a bipartite system from corresponding quantifiers of quantum coherence,
by minimizing the latter over all local bases for one or both subsystems
respectively, has been formalized recently in \cite{ABC}.

\subsection{\label{sub:CV}Coherence in continuous variable systems}

The resource theory framework for quantum coherence adopted in this
Colloquium assumes a finite-dimensional Hilbert space. However, some
of the previously listed quantifiers of coherence have also been studied
in continuous variable systems, and specifically in bosonic modes
of the radiation field. These systems are characterized by an infinite-dimensional
Hilbert space, spanned by the Fock basis $\{\ket{n}\}_{n=0}^{\infty}$
of eigenstates of the particle number operator $a^{\dagger}a$ \cite{BraKimRMP}.

Similarly to what was done by \cite{EisertInfinite} in entanglement theory,
\cite{Zhang2016} imposed a finite mean energy constraint, $\langle a^{\dagger}a\rangle\equiv\bar{n}<\infty$,
to address the quantification of coherence in such systems with respect
to the Fock reference basis. The relative entropy of coherence (see
Sec.~\ref{sub:REC}) was found to maintain its status as a valid
measure of coherence, in particular reaching a finite maximum $C_{r}^{\max}=(\bar{n}+1)\log(\bar{n}+1)-\bar{n}\log{\bar{n}}<\infty$
for any state with finite mean energy $\bar{n}<\infty$. On the
contrary, it was shown that the $l_{1}$-norm of coherence $C_{l_{1}}$
(see Sec.~\ref{sub:l1}) admits no finite maximum and can diverge
even on states with finite  mean energy \cite{Zhang2016}.
This suggests that the $l_{1}$-norm does not provide a suitable quantifier
of coherence in continuous variable systems.

\cite{Xu2015} focused on the quantification of coherence in bosonic
Gaussian states of infinite-dimensional systems, which form an important
subset of states entirely specified by their first and second moments
and useful for theoretical and experimental investigations of continuous
variable quantum information processing \cite{Adesso2007,PirlaRMP}.
In particular, \cite{Xu2015} defined a Gaussian relative entropy
of coherence for a Gaussian state $\varrho$ (with respect to the
Fock basis) as the relative entropy difference between $\varrho$
and the closest incoherent Gaussian state, which is a thermal state
expressible in terms of the first and second moments of $\varrho$.
However, this is only an upper bound on the true relative entropy
of coherence of a Gaussian state $\varrho$, which is still given
by Eq.~(\ref{eq:CrCd}) for any $\varrho$ with finite mean energy
\cite{Zhang2016}, since the closest incoherent state to $\varrho$
given by its diagonal part $\Delta[\varrho]$ in the Fock basis is
not in general a Gaussian state. More recently, \cite{Buono2016}
studied geometric quantifiers of coherence for Gaussian states in
terms of Bures and Hellinger distance from the set of incoherent Gaussian
states (thermal states); once more, these are upper bounds to the
corresponding distance-based coherence monotones as defined in Sec.~\ref{sub:Distance-based-measures}.

Relations between coherence, optical nonclassicality, and entanglement in continuous
variable systems have been investigated recently in \cite{Killoran2015,Vogel2014,VogelScr}, and will be discussed in the next Section.

\subsection{Coherence, asymmetry and nonclassicality\label{sub:Nonclassicality}}

\subsubsection{\label{sub:Asymmetry}Asymmetry monotones}

The relation between coherence and the framework of asymmetry has
been discussed most recently in \cite{Marvian2014a,Marvian2014,Marvian2015,Marvian2016,Piani2016}.
This framework is based on the notion of translationally invariant
operations (TIO), which were already introduced in Sec.~\ref{sub:Classes}.

In \cite{Marvian2015}, the authors proposed postulates for quantifying
the asymmetry of a state with respect to time translations $e^{-iHt}$
induced by a given Hamiltonian $H$. Any asymmetry monotone\footnote{Note that \cite{Marvian2015} call these quantities asymmetry measures. }
$A$ should vanish for all states which are invariant under time
translations, i.e., which are incoherent in the eigenbasis of $H$.
If we denote the latter set by $\mathcal{I}_{H}$, we have
\begin{equation}
A(\varrho)=0\Longleftrightarrow\varrho\in\mathcal{I}_{H}.
\end{equation}
Moreover, $A$ should not increase under translationally invariant
operations $\Lambda\in$ TIO (see Eq.~(\ref{transinv})):
\begin{equation}
A(\Lambda[\varrho])\leq A(\varrho).
\end{equation}
For nondegenerate Hamiltonians, the set IO is strictly larger than
the set TIO \cite{Marvian2015}. It follows that the set of coherence
monotones (intended as IO monotones) is a strict subset of the set of asymmetry monotones (intended as TIO monotones).

An example for an asymmetry monotone which is not a coherence monotone is the \textit{Wigner-Yanase skew information}
\begin{equation}
W(\varrho,H)=-\frac{1}{2}\mathrm{Tr}\left[\left[H,\sqrt{\varrho}\right]^{2}\right],\label{eq:WYI}
\end{equation}
where $[X,Y]=XY-YX$ is the commutator. This quantity was first introduced
by \cite{Wigner1963} as a measure of information, and was proven
to be an asymmetry monotone in \cite{Marvian2012,Marvian2014a,Girolami14}.
While \cite{Girolami14} originally proposed the skew information
as a coherence monotone, it was later shown that such a quantity can instead increase under IO,
in particular under permutations of the reference basis states, hence violating C2 for this set of operations \cite{Du2015,Marvian2015}.

The quantity in Eq.~(\ref{eq:WYI}) can be generalized to the class of Wigner-Yanase-Dyson
skew informations \cite{Wigner1963,Wehrl1978}
\begin{equation}
W_{a}(\varrho,H)=\mathrm{Tr}\left[\varrho H^{2}\right]-\mathrm{Tr}\left[\varrho^{a}H\varrho^{1-a}H\right],
\label{eq:WYD}
\end{equation}
with $0<a<1$, which are also asymmetry monotones \cite{Marvian2012,Marvian2014a}, reducing to Eq.~(\ref{eq:WYI}) for $a=\frac12$. In particular, the average monotone
$\overline{W}(\varrho,H)=\int_{0}^{1} {\rm d} a\  W_{a}(\varrho,H)$,
branded as the {\it quantum variance}, has recently found applications in quantum many-body systems \cite{Frerot2015}.

Another instance of an asymmetry monotone is given by the \textit{quantum
Fisher information} \cite{Braunstein1994,Marvian2014a,Girolami2015,Girolami2016},
which can be defined (under a smoothness hypothesis) as
\begin{equation}
I(\varrho,H)=-2\left.\frac{\partial^{2}F(\varrho_{t},\varrho_{t+\varepsilon})}{\partial\varepsilon^{2}}\right\vert _{\varepsilon\rightarrow0},\label{eq:QFI}
\end{equation}
where $F$ denotes the fidelity defined after Eq.~(\ref{eq:CMG}),
and $\varrho_{t}=e^{-iHt}\varrho e^{iHt}$. The quantum Fisher information
quantifies the sensitivity of the state $\varrho$ to a variation
in the parameter $t$ characterizing a unitary dynamics generated
by $H$, hence playing a central role in quantum metrology \cite{Braunstein1994,Giovannetti2011}.

The Wigner-Yanase-Dyson skew information Eq.~(\ref{eq:WYD}) and the quantum Fisher information Eq.~(\ref{eq:QFI}) can be seen as special instances of the entire family of quantum generalizations of the classical Fisher information \cite{Petz2010}, which are defined in terms of the Riemannian contractive metrics on the quantum state space as classified by \cite{Morozova1991,Petz1996}. All these quantities have been proven to be asymmetry monotones, i.e.~worthwhile quantifiers of unspeakable coherence, in \cite{Girolami2016}.
Further details on the applications of these asymmetry monotones in the contexts of quantum speed limits,
quantum estimation and discrimination \cite{Pires2016,Diego15,Mondal2015,Marvian2015,Marvian2016,Napoli2016,Piani2016},
are reported in Secs.~\ref{sec:speed} and~\ref{sec:metro}.

The concept of asymmetry is closely related to the concept of quantum
reference frames \cite{Bartlett2007,Vaccaro2008,Gour2008,Gour2009}.
Initially, these two concepts were defined for an arbitrary Lie group
$G$. If $U(g)$ is a unitary representation of the group with $g\in G$,
the set of invariant states ${\cal G}$ consists of all states
which are invariant under the action of all $U(g)$. The \emph{relative
entropy of asymmetry} (also known as \emph{relative entropy of
frameness}) is then defined as \cite{Gour2009}
\begin{equation}
A_{r}(\varrho)=\min_{\sigma\in {\cal G}}S(\varrho||\sigma).
\end{equation}
Remarkably, it was shown by \cite{Gour2009} that the relative entropy
of asymmetry admits the following expression, first independently introduced as $G$-asymmetry by \cite{Vaccaro2008}:
\begin{equation}\label{eq:G-asy}
A_{r}(\varrho)=S\left(\Gamma\left[\varrho\right]\right)-S(\varrho)=S\left(\varrho||\Gamma\left[\varrho\right]\right),
\end{equation}
where $\Gamma[\varrho]=\int {\rm d}g U(g)\varrho U^{\dagger}(g)$ is
the average with respect to the Haar measure $dg$. If the unitary
representation of the group is given by $\left\{ e^{-iHt}:\,t\in\mathbb{R}\right\} $
with a Hermitian nondegenerate matrix $H=\sum_{i}h_{i}\ket{i}\!\bra{i}$,
the set ${\cal G}$ of invariant states  is precisely the set ${\cal I}$ of
states which are incoherent in the eigenbasis $\{\ket{i}\}$ of $H$.
Thus, in this case the relative entropy of asymmetry $A_{r}$ is equal
to the relative entropy of coherence $C_{r}$ with respect to the
eigenbasis of $H$ taken as a reference basis.

The {\it robustness of asymmetry} with respect to an arbitrary Lie group $G$ is an asymmetry monotone defined in \cite{Piani2016,Napoli2016} as
\begin{equation}
R_{A}(\varrho)=\min_{\tau}\left\{ s\geq0\left|\frac{\varrho+s\tau}{1+s}\in {\cal G}\right.\right\} .\label{eq:RobustnessA}
\end{equation}
If the unitary representation of the group is given by $e^{-iHt}$ as above, then the robustness of asymmetry $R_A$ reduces to the robustness of coherence $R_C$ with respect to the eigenbasis of $H$, defined in Eq.~(\ref{eq:Robustness}).

\subsubsection{Quantifying superpositions}
A very general approach to quantify coherence was presented by \cite{Aberg2006}
within the framework of \emph{quantum superposition}. In this approach,
the Hilbert space $\mathcal{H}$ is divided into $K$ subspaces ${\cal L}_{1},\ldots,{\cal L}_{K}$
such that $\oplus_{k=1}^{K}\mathcal{L}_{k}=\mathcal{H}$. If $P_{k}$
is the projector corresponding to the subspace $\mathcal{L}_{k}$,
the total operation $\Pi$ is defined as
\begin{equation}
\Pi[\varrho]=\sum_{k=1}^{K}P_{k}\varrho P_{k}.
\end{equation}
If the projectors $P_{k}$ all have rank one, the total operation
$\Pi$ corresponds to the total dephasing $\Delta$. However, in general
the projectors $P_{k}$ can have rank larger than one.

\cite{Aberg2006} also proposed a set of conditions a faithful quantifier
of superposition ought to satisfy and showed that the relative entropy
of superposition fulfills these conditions. The latter is defined
as follows:
\begin{equation}
S_{r}(\varrho)=S(\Pi[\varrho])-S(\varrho).
\end{equation}
The relative entropy of superposition is a special case of the relative
entropy of asymmetry presented in Sec.~\ref{sub:Asymmetry}, and
admits the following expression \cite{Aberg2006}:
\begin{equation}
S_{r}(\varrho)=\min_{\Pi[\sigma]=\sigma}S(\varrho||\sigma)=S(\varrho||\Pi[\varrho]).
\end{equation}

\subsubsection{\label{sub:General-measures-of-nonclassicality}Coherence rank and general quantifiers
of nonclassicality}

An alternative approach was taken by \cite{Killoran2015,Theurer2017,Mukhopadhyay2017,Regula2017}, who investigated
a very general form of nonclassicality, also going beyond the framework
of coherence. In particular, depending on the task under consideration,
it can be useful to identify a set of pure states $\{\ket{c_{i}}\}$
as classical. These states do not have to be mutually orthogonal in
general. As an example, in entanglement theory those are all states
of the product form $\ket{c}=\ket{\alpha}\otimes\ket{\beta}$.

\cite{Killoran2015} introduced the \emph{coherence rank} of a general
pure state\footnote{The coherence rank can be generalized to mixed states via a procedure similar to the convex roof described in Sec.~\ref{sub:Convex-roof-measures}. The resulting quantity $r_C(\varrho)= \inf_{\{p_{i},\ket{\psi_{i}}\}} \max_i r_C(\ket{\psi_i})$ is called the {\it coherence number} of $\varrho$ \cite{Regula2017}.}.
Analogously to the Schmidt rank in entanglement theory, the coherence
rank counts the minimal number of classical states needed in the expansion
of the general state $\ket{\psi}$:
\begin{equation}\label{eq:rank}
r_{C}(\ket{\psi})=\min\left\{ R:\ket{\psi}=\sum_{i=0}^{R-1}a_{i}\ket{c_{i}}\right\} .
\end{equation}
The authors of \cite{Killoran2015} then proved that nonclassicality
can always be converted into entanglement in the following sense:
if the set of classical states $\{\ket{c_{i}}\}$ is linearly independent,
there always exists a unitary operation which converts each state
$\ket{\psi}$ with coherence rank $r_{C}$ into a bipartite state
$\ket{\tilde{\psi}}$ with the same Schmidt rank. The authors of \cite{Regula2017} extended this result to the multipartite setting, by constructing a unitary protocol for converting nonclassicality of a $d$-level system, prepared in any input state with coherence rank $2 \leq r_C \leq d$, into genuine $(r_C+1)$-partite entanglement between the system and up to $d$ ancillary qubits.

A concept related to the coherence rank was discussed in \cite{Levi14} in the specific context of  \emph{coherent
delocalization}. In this framework, a state $\ket{\psi}$ is called
$k$-coherent if it can be written as $\ket{\psi}=\sum_{i=0}^{k-1}a_{i}\ket{i}$
with all coefficients $a_{i}$ being nonzero. Here the integer $k$ corresponds to the coherence rank $r_C(\ket{\psi})$, where the set of classical states $\{\ket{c}_i\}$ in the definition of Eq.~(\ref{eq:rank}) is identified with the reference basis of incoherent states $\{\ket{i}\}$.
\cite{Levi14} also
proposed quantifiers for this concept of $k$-coherence, and showed that
these quantifiers do not grow under incoherent channels in their framework.
Note however that the incoherent channels of \cite{Levi14} are in
general different from the IO defined by \cite{Baumgratz2014}, and
can be rather identified with the SIO \cite{Winter2015,Yadin2015b}.
We also note that a related framework was presented recently by \cite{Yadin2015}
to quantify macroscopic coherence. The possibility to establish superpositions of unknown quantum states via universal quantum protocols has been investigated by \cite{Oszmaniec2016}.

\subsubsection{Optical coherence and nonclassicality}
The framework of \cite{Killoran2015,Theurer2017} is partly motivated by the seminal theory of optical coherence in continuous variable systems \cite{Sudarshan1963,Glauber1963,Mandel1965}. In this theory, the pure classical states are identified with the Glauber-Sudarshan coherent states $\ket{\alpha}$ of the radiation field, defined (for a single bosonic mode) as the right eigenstates of the annihilation operator, $a \ket{\alpha} = \alpha \ket{\alpha}$, with $\alpha \in \mathbb{C}^2$. These states form an overcomplete, non-orthogonal basis for the infinite-dimensional Hilbert space. Any quantum state $\varrho$ which cannot be written as a mixture of Glauber-Sudarshan coherent states is hence regarded as nonclassical. We wish to highlight a semantic subtlety here: a Glauber-Sudarshan coherent state $\ket{\alpha}$ (with $\alpha \neq 0$) is in fact coherent if one is interested in characterizing coherence with respect to the Fock basis $\{\ket{n}\}$ as in Sec.~\ref{sub:CV}, since it can be written as a superposition thereof: $\ket{\alpha} = e^{-|\alpha|^2/2}\sum_{n=0}^\infty \frac{\alpha^n}{\sqrt{n!}} \ket{n}$. However, in the theory of optical coherence the Glauber-Sudarshan coherent states play rather the role of classical, or free, states (i.e., the analogous of incoherent states in the resource theory of coherence discussed so far), as they can be generated by classical currents acting
on a quantum field \cite{Louisell1973}. Hence it is well motivated that they form the reference set with respect to which
the resource of {\it optical nonclassicality} is defined and quantified.

A general resource theory of nonclassicality, as  suggested in \cite{Brandao2008}, has not been completely formalized yet. In particular, determining the suitable set of free operations for the theory of optical coherence stands as one of the most pressing open questions. The set ${\cal CO}$ of {\it classical operations}, defined as the maximal set of operations preserving a reference set of (not mutually orthogonal) classical states, has been studied in \cite{VogelScr}, where it was shown that ${\cal CO}$ is convex and obeys the semigroup property. If the set of classical states is identified with the convex hull of Glauber-Sudarshan coherent states, as in the theory of optical coherence, then the corresponding ${\cal CO}$ includes so-called {\it passive operations}, i.e.~operations preserving the mean energy $\langle a^\dagger a\rangle$, which can be implemented by linear optical elements such as beam splitters and phase shifters.

A series of works investigated the conversion from optical nonclassicality into entanglement by means of passive operations \cite{Kim02,Wolf2003,Asboth05,Vogel2014,VogelScr}, serving as an inspiration for the more recent studies of \cite{Streltsov2015b} and \cite{Killoran2015,Theurer2017} reviewed in the previous Sections.

In particular, \cite{Asboth05} proposed to quantify optical nonclassicality for a single-mode state $\varrho$ in terms of the maximum two-mode entanglement that can be generated from $\varrho$ using linear optics, auxiliary classical states, and ideal photodetectors. Any output entanglement monotone $E$ defines a corresponding nonclassicality quantifier $P_E$ for the input state $\varrho$, referred to as {\it entanglement potential}. The authors of \cite{Asboth05} considered in particular the quantities $P_E$ derived by choosing $E$ to be the logarithmic negativity or the relative entropy of entanglement. The definition of entanglement-based coherence monotones by \cite{Streltsov2015b} as presented in Sec.~\ref{sub:Coherence-Entanglement} can be seen as the finite-dimensional counterpart to the study of \cite{Asboth05}.

Furthermore, \cite{Vogel2014} independently defined a notion analogous to the coherence rank $r_C$ of Eq.~(\ref{eq:rank}) for optical nonclassicality, i.e.~with $\{\ket{c_{i}}\}$ being a subset of (linearly independent) Glauber-Sudarshan coherent states. They then showed that a single-mode state $\ket{\psi}$ with nonclassicality rank $r_C$ can always be mapped into a two-mode entangled state with the same Schmidt rank, by means of a balanced beam splitter acting on the input mode and a vacuum ancillary mode. This can be seen as a special instance of the general theorem of \cite{Killoran2015} presented in Sec.~\ref{sub:General-measures-of-nonclassicality}.

Finally, a connection between optical nonclassicality and the theory of coherence as reviewed in Sec.~\ref{sec:resource-theories} with respect to the set IO has been recently established by \cite{Jeong2017}. They introduced an orthogonalization procedure, according to which one can define quantifiers of optical nonclassicality $P_C$, i.e.~coherence with respect to the non-orthogonal basis of Glauber-Sudarshan coherent states, in terms of any coherence monotone $C$ applied to $N$-dimensional subspaces of the Hilbert space, in a suitable limit $N \rightarrow \infty$; details of the mapping can be found in \cite{Jeong2017}. They then proved that any such $P_C$ is a monotone under linear optical passive operations if the corresponding $C$ is an IO monotone.
This demonstrates that continuous variable states exhibiting optical nonclassicality can be seen essentially as the limiting case of the same resource states identified in \cite{Baumgratz2014},  when the incoherent basis is chosen as the set of Glauber-Sudarshan coherent states.

\subsection{Multipartite settings}

\subsubsection{General distance-based coherence quantifiers}

In the bipartite setting it is possible to obtain coherence monotones
$C_{D}$ by using the distance-based approach as in Eq.~(\ref{eq:CDistance}),
where ${\cal I}$ is now the set of bipartite incoherent states, i.e.,
convex combinations of states of the form $\ket{k}\ket{l}$, with
$\{\ket{k}\}$ and $\{\ket{l}\}$ the incoherent reference bases for
each subsystem respectively \cite{Bromley2014,Streltsov2015b}. It
is instrumental to compare the quantities obtained in this way to
other corresponding distance-based quantifiers of bipartite nonclassicality
such as quantumness $Q_{D}$ and entanglement $E_{D}$. Due to the
inclusion relation ${\cal I}\subset{\cal CC}\subset{\cal S}$, the
aforementioned quantities are related via the following inequality
\cite{Yao2015}:
\begin{equation}
C_{D}(\varrho)\geq Q_{D}(\varrho)\geq E_{D}(\varrho).
\end{equation}
These results can be straightforwardly generalized to more than two
parties \cite{Yao2015}.

\subsubsection{Quantum-incoherent relative entropy}

The {\em quantum-incoherent relative entropy} was defined by \cite{Chitambar2015}
as follows:
\begin{equation}
C_{r}^{A|B}\left(\varrho^{AB}\right)=\min_{\sigma^{AB}\in\mathcal{QI}}S\left(\varrho^{AB}||\sigma^{AB}\right),
\end{equation}
where the minimum is taken over the set of quantum-incoherent states
$\mathcal{QI}$ defined in Eq.~(\ref{eq:qi}). As further discussed
in \cite{Chitambar2015}, the quantum-incoherent relative entropy
admits the following closed expression:
\begin{equation}
C_{r}^{A|B}\left(\varrho^{AB}\right)=S\left(\Delta^{B}\left[\varrho^{AB}\right]\right)-S\left(\varrho^{AB}\right),
\end{equation}
where $\Delta^{B}$ denotes a dephasing operation on subsystem $B$
only. As we will see in the following, the quantum-incoherent relative
entropy is a powerful upper bound on the distillable coherence of
collaboration.

\subsubsection{\label{sub:Distillable-coherence-of-collaboration}Distillable coherence
of collaboration}

The {\em distillable coherence of collaboration} was introduced
and studied by \cite{Chitambar2015} as the figure of merit for the
task of assisted coherence distillation. In this task Alice and Bob
share a bipartite state $\varrho^{AB}$ and aim to extract maximally
coherent single-qubit states $\ket{\Psi_{2}}$ on Bob's side via LQICC
operations. The distillable coherence of collaboration is the highest
achievable rate for this procedure \cite{Chitambar2015}:
\begin{eqnarray}
C_{\mathrm{LQICC}}^{A|B}(\varrho) & = & \sup\left\{ R:\lim_{n\rightarrow\infty}\left(\inf_{\Lambda}\left\Vert \mathrm{Tr}_{A}\left[\Lambda\left[\varrho^{\otimes n}\right]\right]-{\tau}^{\otimes\left\lfloor Rn\right\rfloor }\right\Vert _{1}\right)=0\right\} .\nonumber \\
\label{eq:coherence-of-collaboration}
\end{eqnarray}
Here, the infimum is taken over all LQICC operations $\Lambda$, and
$\tau=\ket{\Psi_{2}}\!\bra{\Psi_{2}}^{B}$ is the maximally coherent
single-qubit state on Bob's subsystem. As was shown in \cite{Chitambar2015},
for a pure state $\ket{\psi}^{AB}$ the distillable coherence of collaboration
is equal to the regularized coherence of assistance of Bob's reduced
state:
\begin{equation}
C_{\mathrm{LQICC}}^{A|B}(\psi^{AB})=C_{a}^{\infty}(\varrho^{B})=S(\Delta[\varrho^{B}]).\label{eq:collaboration}
\end{equation}
It is interesting to compare this result to the distillable coherence
of Bob's local state: $C_{d}(\varrho^{B})=S(\Delta^{B}[\varrho^{B}])-S(\varrho^{B})$.
This means that assistance provides an improvement on the distillation
rate given exactly by the local von Neumann entropy $S(\varrho^{B})$.
Remarkably, this improvement does not depend on the particular choice
of the incoherent reference basis.

In \cite{Streltsov2015c} this framework was extended to other sets
of operations, such as LICC, SI, and SQI, see Sec.~\ref{sub:LICC}
for their definitions. In general, if $X$ is one of the sets described
above, then the distillable coherence of collaboration can be generalized
as follows \cite{Streltsov2015c}:
\begin{equation}
C_{X}^{A|B}(\varrho)=\sup\left\{ R:\lim_{n\rightarrow\infty}\left(\inf_{\Lambda\in X}\left\Vert \mathrm{Tr}_{A}\left[\Lambda\left[\varrho^{\otimes n}\right]\right]-{\tau}^{\otimes\left\lfloor Rn\right\rfloor }\right\Vert _{1}\right)=0\right\} .
\end{equation}
Interestingly, for any mixed state $\varrho=\varrho^{AB}$ the quantities
$C_{\mathrm{SI}}^{A|B}$ and $C_{\mathrm{SQI}}^{A|B}$ are equal,
and all quantities $C_{X}^{A|B}$ are between $C_{\mathrm{LICC}}^{A|B}$
and $C_{r}^{A|B}$ \cite{Streltsov2015c}:
\begin{equation}
C_{\mathrm{LICC}}^{A|B}\leq C_{\mathrm{LQICC}}^{A|B}\leq C_{\mathrm{SI}}^{A|B}=C_{\mathrm{SQI}}^{A|B}\leq C_{r}^{A|B}.
\end{equation}
Moreover, for bipartite pure states $\ket{\psi}^{AB}$, Eq.~(\ref{eq:collaboration})
generalizes as follows \cite{Streltsov2015c}:
\begin{align}
C_{\mathrm{LICC}}^{A|B}(\ket{\psi}^{AB}) & =C_{\mathrm{LQICC}}^{A|B}(\ket{\psi}^{AB})=C_{\mathrm{SI}}^{A|B}(\ket{\psi}^{AB})=C_{\mathrm{SQI}}^{A|B}(\ket{\psi}^{AB})\nonumber \\
 & =C_{r}^{A|B}(\ket{\psi}^{AB})=C_{a}^{\infty}(\varrho^{B})=S(\Delta[\varrho^{B}]).
\end{align}

Very recently \cite{Wu2017} provided the first experimental
demonstration of assisted coherence distillation. The experiment
used polarization-entangled photon pairs for creating pure entangled
states, and also mixed Werner states. After performing a suitable
measurement on one of the photons,  the second photon was found in
a state with a larger amount of coherence.

Finally, we note that the distillable coherence of collaboration can
be regarded as the coherence analogue of the entanglement of collaboration
presented in \cite{Gour2006}.

\subsubsection{\label{sub:Recoverable-coherence}Recoverable coherence}

The {\em recoverable coherence} was introduced by \cite{Matera2015}
in the context of a resource theory of control of quantum systems.
It is defined in the same way as the distillable coherence of collaboration
in Eq.~(\ref{eq:coherence-of-collaboration}), but with the set of
LQICC operations replaced by GOI$B$, see Sec.~\ref{sub:LICC} for
their definition. Following the analogy to distillable coherence of
collaboration, we will denote the recoverable coherence by $C_{\mathrm{GOI}B}^{A|B}$.
As was shown by \cite{Matera2015}, the recoverable coherence is additive,
convex, monotonic on average under GOI$B$ operations, and upper bounded
by the quantum-incoherent relative entropy. Since LQICC is a subset
of GOI$B$ operations, we get the following inequality:
\begin{equation}
C_{\mathrm{LQICC}}^{A|B}\leq C_{\mathrm{GOI}B}^{A|B}\leq C_{r}^{A|B}.
\end{equation}
Notably, the recoverable coherence has an operational interpretation,
as it is directly related to the precision of estimating the trace
of a unitary via the DQC1 quantum algorithm \cite{Matera2015}; see
also Sec.~\ref{sub:Quantum-algorithms} for a more general discussion
on the role of coherence in quantum algorithms. Additionally, minimizing
the recoverable coherence over all local bases leads to an alternative
quantifier of discord \cite{Matera2015}.

\subsubsection{Uncertainty relations and monogamy of coherence }

Uncertainty relations for quantum coherence, both for a single party
and for multipartite settings, have been studied in \cite{Singh2016,Peng2016b}.
If coherence is defined with respect to two different bases, $\{\ket{i}\}$
and $\{\ket{a}\}$, the corresponding relative entropies of coherence
$C_{r}^{i}$ and $C_{r}^{a}$ fulfill the following uncertainty relation
\cite{Singh2016}:
\begin{equation}
C_{r}^{i}(\varrho)+C_{r}^{a}(\varrho)\geq-2\log_{2}\left(\max_{i,a}|\!\braket{i|a}\!|\right)-S(\varrho).
\end{equation}
For bipartite states $\varrho^{XY}$, \cite{Singh2016} derived the following
uncertainty relation for the bipartite relative entropies of coherence
$C_{r}^{ij}$ and $C_{r}^{ab}$:
\begin{equation}
C_{r}^{ij}(\varrho^{XY})+C_{r}^{ab}(\varrho^{XY})\leq2\log_{2}d_{XY}-2K(\varrho^{XY}).
\end{equation}
Here, $d_{XY}$ is the dimension of the composite Hilbert space, and $K(\varrho^{XY})$
arises from the Lewenstein-Sanpera decomposition\footnote{{A Lewenstein-Sanpera decomposition \cite{Lewenstein1998} of
a state $\varrho$ is a decomposition of the form $\varrho=\lambda\varrho_{s}+(1-\lambda)\varrho_{e}$
with a separable state $\varrho_{s}$, an entangled state $\varrho_{e}$,
and probability $0\leq\lambda\leq1$.}}: $K(\varrho^{XY})=\lambda S(\varrho_{s}^{XY})+(1-\lambda)S(\varrho_{e}^{XY})$.

The discussion on {\it monogamy} of quantum coherence is also inspired by
results from entanglement theory \cite{coffman2000distributed,HorodeckiRMP09}.
In particular, a coherence quantifier $C$ is called monogamous with
respect to the subsystem $X$ for a tripartite state $\varrho^{XYZ}$
if \cite{Yao2015,Kumar2015}
\begin{equation}
C(\varrho^{XYZ})\geq C(\varrho^{XY})+C(\varrho^{XZ}).
\end{equation}
As was shown in \cite{Yao2015,Kumar2015}, the relative entropy of
coherence is not monogamous in general, although it can be monogamous
for certain families of states. Further results on monogamy of coherence
have also been presented by \cite{Radhakrishnan2016}.

\section{\label{sec:dynamics}Dynamics of quantum coherence}

Quantum coherence is typically recognized as a fragile feature: the
vanishing of coherence in open quantum systems exposed to environmental
noise, commonly referred to as \textit{decoherence} \cite{Breuer2002,Zurek2003,Schlosshauer2005},
is perhaps the most distinctive manifestation of the quantum-to-classical
transition observed at our macroscopic scales. Numerous efforts have
been invested into devising feasible control schemes to preserve coherence
in open quantum systems, with notable examples including dynamical
decoupling \cite{DynDec}, quantum feedback control \cite{FeedControl}
and error correcting codes \cite{Shor1995}.

In this Section we review more recent work concerning the dynamical
evolution of coherence quantifiers (defined in Sec.~\ref{sec:Quantifying-coherence})
subject to relevant Markovian or non-Markovian evolutions. Coherence
effects in biological systems and their potential functional role
will be discussed in Sec.~\ref{sec:applications}. Here we also discuss
generic properties of coherence in mixed quantum states, the cohering
(and decohering) power of quantum channels, and the role played by
coherence quantifiers in defining speed limits for closed and open
quantum evolutions.

\subsection{Freezing of coherence}

\label{Sec:4.Freezing} One of the most interesting phenomena observed
in the dynamics of coherence is the possibility for its \textit{freezing},
that is, complete time invariance without any external control, in
the presence of particular initial states and noisy evolutions. \cite{Bromley2014}
identified a set of dynamical conditions under which \textit{all}
distance-based coherence monotones $C_{D}$ obeying postulates C1,
C2 (for the set IO), and C4 stay simultaneously frozen for indefinite
time (see also Fig.~\ref{fig:plan}, bottom panel). We can summarize
these conditions as follows.

Let us consider an open quantum system of $N$ qubits, each subject
to a nondissipative Markovian decoherence channel, representing dephasing
the eigenbasis of the $k$th Pauli matrix $\sigma_{k}$, where $k=1$
corresponds to bit flip noise, $k=2$ to bit-phase flip noise, and
$k=3$ to phase flip noise (the latter equivalent to conventional
phase damping in the computational basis). Such `$k$-flip' channels
on each qubit are described by a set of Kraus operators \cite{Nielsen10}
$K_{0}(t)=\sqrt{1-q(t)/2}\ \openone$, $K_{i,j\neq k}(t)=0$, $K_{k}(t)=\sqrt{q(t)/2}\ \sigma_{k}$,
where $\{i,j,k\}$ is a permutation of $\{1,2,3\}$ and $q(t)=1-e^{-\gamma t}$
is the strength of the noise, with $\gamma$ the decoherence rate.
Any such dynamics, mapping a $N$-qubit state $\varrho(0)$ into $\varrho(t)=\sum_{m_{1},\ldots,m_{N}=0}^{3}\big(K_{m_{1}}(t)\otimes\ldots\otimes K_{m_{N}}(t)\big)\varrho(0)\big(K_{m_{1}}(t)\otimes\ldots\otimes K_{m_{N}}(t)\big)^{\dagger}$,
is incoherent (in particular, strictly incoherent) with respect to
any product basis $\{\ket{m}\}^{\otimes N}$, with $\{\ket{m}\}$
being the eigenbasis of any of the three canonical Pauli operators
$\sigma_{m}$ on each qubit.

Let us now consider a family of $N$-qubit mixed states with all maximally
mixed marginals, defined by
\begin{equation}
\varrho(t)=\frac{1}{2^{N}}\left(\openone^{\otimes N}+\sum_{j=1}^{3}c_{j}(t)\sigma_{j}^{\otimes N}\right)\,,\label{eq:M3N}
\end{equation}
with $c_{j}=\mathrm{Tr}[\varrho\sigma_{j}^{\otimes N}]$. For any
even $N$, these states span a tetrahedron with vertices $\{1,(-1)^{N/2},1\}$,
$\{-1,-(-1)^{N/2},1\}$, $\{1,-(-1)^{N/2},-1\}$, and $\{-1,(-1)^{N/2},-1\}$,
in the three-dimensional space of the correlation functions $\{c_{1},c_{2},c_{3}\}$.
In the special case $N=2$, they reduce to Bell diagonal states of
two qubits, that is, arbitrary convex mixtures of the four maximally
entangled Bell states. Freezing of coherence under $k$-flip channels
then manifests for the subclass of states of Eq.~(\ref{eq:M3N})
with initial condition $c_{i}(0)=(-1)^{N/2}c_{j}(0)c_{k}(0)$, for
any even $N\geq2$. For all such states, and for any distance functional
$D(\varrho,\sigma)$ being zero for $\sigma=\varrho$, contractive
under quantum channels, and jointly convex, one has
\begin{equation}
C_{D}\big(\varrho(t)\big)=C_{D}\big(\varrho(0)\big)\quad\forall\,t\geq0\,,\label{eq:Frozen}
\end{equation}
where $C_{D}$ is a corresponding distance-based quantifier of coherence
(see Sec.~\ref{sub:Distance-based-measures}), and coherence is measured
with respect to the product eigenbasis $\{\ket{j}\}^{\otimes N}$
of the Pauli operator $\sigma_{j}^{\otimes N}$ \cite{Bromley2014,Silva2015}.
Notice that the freezing extends as well to the $l_{1}$-norm of coherence,
as it amounts to the trace distance of coherence in the considered
states (up to a normalization factor). For odd $N$, including the
general case of a single qubit, no measure-independent freezing of
coherence can occur instead for the states of Eq.~(\ref{eq:M3N}),
apart from trivial instances; this means that, for all nontrivial
evolutions preserving e.g.~the $l_{1}$-norm of coherence $C_{l_{1}}$,
some other quantifier such as the relative entropy of coherence $C_{r}$
is strictly decreasing \cite{Bromley2014}.

In \cite{Yu2016}, the relative entropy of coherence was found to
play in fact a special role in identifying conditions such that any
coherence monotone is frozen at all times. Specifically, all coherence
monotones (respecting in particular property C2 for the set SIO) are
frozen for an initial state subject to a strictly incoherent channel
if and only if the relative entropy of coherence is frozen for such
initial state \cite{Yu2016}. In formulae,
\begin{equation}
C\big(\varrho(t)\big)=C\big(\varrho(0)\big)\quad\forall\,C\quad\Leftrightarrow\quad C_{r}\big(\varrho(t)\big)=C_{r}\big(\varrho(0)\big)\,,\label{eq:FreezingYu}
\end{equation}
where $\varrho(t)=\Lambda_{t}[\varrho(0)]$ with $\Lambda_{t}\in$
SIO (see Sec.~\ref{sub:Classes}). Using this criterion, one can
identify other classes of initial states exhibiting measure-independent
frozen coherence under local $k$-flip channels \cite{Yu2016}.

We remark that the described freezing effect differs from an instance
of \textit{decoherence-free subspace} \cite{Lidar2014}, where an
open system dynamics acts effectively as a unitary evolution on a
subset of quantum states, preserving their informational properties.
Here, instead, the purity of the involved states is degraded with
time, but their coherence in the chosen reference basis remains unaffected.
Other signatures of quantumness like measures of discord-type correlations
can also freeze under the same dynamical conditions \cite{Modi2012,Mazzola2010,Cianciaruso2015},
albeit only for a finite time in the case of Markovian dynamics. A
unified geometric analysis of these phenomena is given in \cite{Bromley2014,Cianciaruso2015,Silva2015}.

In \cite{Liu2015}, freezing of coherence was explored theoretically
for a system of two-level atoms interacting with the vacuum fluctuations
of an electromagnetic field bath. A more comprehensive analysis of
the dynamics of the $l_{1}$-norm of coherence for one qubit subject
to various types of common noisy channels was reported in \cite{Maziero2016}.
The dynamics of the $l_{1}$-norm of coherence for general $d$-dimensional systems was further investigated in \cite{Hu2016},
where a factorization relation for the evolution equation of $C_{l_{1}}$ was derived, leading in particular to a condition for its freezing.

The phenomenon of frozen quantum coherence was demonstrated experimentally
in a room temperature nuclear magnetic resonance setup \cite{Silva2015},
with two-qubit and four-qubit ensembles prepared in states of the
form~(\ref{eq:M3N}).

\subsection{Coherence in non-Markovian evolutions}

Some attention has been devoted to the study of coherence in non-Markovian
dynamics. In \cite{Addis2014}, the phenomenon of \textit{coherence
trapping} in the presence of non-Markovian dephasing was studied.
Namely, for a single qubit subject to non-Markovian pure dephasing
evolutions (i.e., a $k$-flip channel with $\gamma t$ replaced by
a non-monotonic function of $t$), the stationary state at $t\rightarrow\infty$
may retain a nonzero coherence in the eigenbasis of $\sigma_{k}$,
as quantified e.g.~by the $l_{1}$-norm of coherence. This can only
occur in the presence of non-Markovian dynamics, and is different
from the previously discussed case of coherence freezing, in which
coherence is measured instead with respect to a reference basis transversal
to the dephasing direction. It was shown in \cite{Addis2014} that
the specifics of coherence trapping depend on the environmental spectrum:
its low-frequency band determines the presence or absence of information
backflow, while its high-frequency band determines the maximum coherence
trapped in the stationary state. In \cite{Zhang2015} the coherence
in the stationary state of a qubit initially correlated with a zero-temperature
Ohmic-like bath, realizing a non-Markovian pure dephasing channel,
was further studied. The best dynamical conditions were identified
(in terms of initial qubit-bath correlations and bath spectral density)
to optimize coherence trapping, that is, to maximize coherence in
the stationary qubit state and to minimize the evolution time towards
such a state.

The dynamics of the $l_{1}$-norm of coherence for two qubits globally
interacting with a harmonic oscillator bath was investigated in \cite{Bhattacharya2016},
finding that non-Markovianity slows down the coherence decay. A proposal
to witness non-Markovianity of incoherent evolutions via a temporary
increase of coherence quantifiers was finally discussed in \cite{Chanda2015},
inspired by more general approaches to witness and measure non-Markovianity
based on revivals of distinguishability, entanglement, or other informational
quantifiers \cite{Rivas2014R,Breuer2016R}.

\subsection{\label{sub:coheringpower}Cohering power of quantum channels and
evolutions}

The \textit{cohering power} of a quantum channel \cite{Baumgratz2014,Mani2015}
can be defined as the maximum amount of coherence that the channel
can create when acting on an incoherent state,
\begin{equation}
{\cal P}_{C}(\Lambda)=\max_{\varrho\in\mathcal{I}}C(\Lambda[\varrho])\,,\label{eq:CoheringPower}
\end{equation}
where $\mathcal{I}$ denotes the set of incoherent states (with respect
to a chosen reference basis) and $C$ is any quantifier of coherence.
In a similar way, \cite{Mani2015} defined the \textit{decohering
power} of a quantum channel $\Lambda$ as the maximum amount by which
the channel reduces the coherence of a maximally coherent state,
\begin{equation}
{\cal D}_{C}(\Lambda)=\max_{\varrho\in\mathcal{M}}\left\{ C(\varrho)-C(\Lambda[\varrho])\right\} \,,\label{eq:DecoheringPower}
\end{equation}
where $\mathcal{M}$ denotes the set of maximally coherent states
(with respect to a chosen reference basis) and $C$ is any quantifier
of coherence. If $C$ is chosen to be convex (i.e., respecting property
C4), then the optimizations in Eqs.~(\ref{eq:CoheringPower}) and
(\ref{eq:DecoheringPower}) are achieved by pure (respectively incoherent
and maximally coherent) states, simplifying the evaluation of the
cohering and decohering power of a quantum channel.

Refs.~\cite{Mani2015,Bu2015,Xi2015,Garcia-Diaz2015,Situ2016,Bu2016} calculated
the cohering and decohering power of various unitary and non-unitary
quantum channels adopting different quantifiers of coherence. In particular,
\cite{Mani2015} showed that ${\cal P}_{C}(\Lambda)={\cal D}_{C}(\Lambda)$
for all single-qubit unitary channels when adopting the Wigner-Yanase
skew information as a quantifier of coherence $C$.\footnote{More precisely,
this would quantify the asymmetry/symmetry power rather than the cohering/decohering
power.} In \cite{Bu2015}, the authors derived a closed expression
for the cohering power of a unitary channel $U$ when adopting the
$l_{1}$-norm as a quantifier of coherence,
\begin{equation}
{\cal P}_{C_{l_{1}}}(U)=\|U\|_{1 \rightarrow 1}^{2}-1\,,\label{eq:CoheringPowerU}
\end{equation}
where $\|U\|_{1 \rightarrow 1} = \max_{\|x\|_1=1} \|U x\|_1$ denotes the maximum column sum matrix norm. This implies
that, for the cohering power of a tensor product of unitaries $\bigotimes_{j}U_{j}$
with respect to a product basis, one has ${\cal P}_{C_{l_{1}}}\left(\bigotimes_{j}U_{j}\right)+1=\prod_{j}\left[{\cal P}_{C_{l_{1}}}(U_{j})+1\right]$,
which generalizes an expression already obtained in \cite{Mani2015}
in the case of all equal $U_{j}$. In \cite{Bu2015} it was also proven
that the $N$-qubit unitary operation with the maximal $l_{1}$-norm
cohering power (even including arbitrary global unitaries) is the
tensor product $H^{\otimes N}$ of $N$ single-qubit Hadamard gates,
with ${\cal P}_{C_{l_{1}}}(H^{\otimes N})=2^{N}-1$. Further examples
of cohering and decohering power of quantum channels with respect
to $l_{1}$-norm, relative entropy, and other coherence quantifiers
were presented in \cite{Bu2015,Xi2015,Garcia-Diaz2015,Situ2016}.

The authors of \cite{Bu2015} provided an operational interpretation
for the cohering power. Given a quantum channel $\Lambda:\mathcal{B}(\mathcal{H})\rightarrow\mathcal{B}(\mathcal{H})$
acting on a principal system, it is said to be implementable by incoherent
operations supplemented by an ancillary quantum system if there exists
an incoherent operation IO $\ni\Lambda_{\mathrm{i}}:\mathcal{B}(\mathcal{H}\otimes\mathcal{H}')\rightarrow\mathcal{B}(\mathcal{H}\otimes\mathcal{H}')$
and states $\sigma,\sigma'\in\mathcal{B}(\mathcal{H}')$ of the ancilla
such that, for any state $\varrho\in\mathcal{B}(\mathcal{H})$ of
the principal system, one has $\Lambda_{\mathrm{i}}[\varrho\otimes\sigma]=\Lambda[\varrho]\otimes\sigma'$.
In this setting, the cohering power of $\Lambda$ quantifies the minimum
amount of coherence to be supplied in the ancillary state $\sigma$
to make $\Lambda$ implementable by incoherent operations: $C(\sigma)\geq{\cal P}_{C}(\Lambda)$,
where $C$ is any coherence monotone fulfilling C1--C4.

On the other hand, \cite{Bu2015,Garcia-Diaz2015} considered a more
general definition of cohering power of a quantum channel $\Lambda$,
given by the maximum increase of coherence resulting from the action
of the channel on an arbitrary state,
\begin{equation}
\widetilde{{\cal P}}_{C}(\Lambda)=\max_{\varrho}\left\{ C(\Lambda[\varrho])-C(\varrho)\right\} \,,\label{eq:PCoheringPower}
\end{equation}
where $C$ is any quantifier of coherence and, unlike Eq.~(\ref{eq:CoheringPower}),
$\varrho$ is not restricted to be an incoherent state. By definition,
${\cal P}_{C}(\Lambda)\leq\widetilde{{\cal P}}_{C}(\Lambda)$. When
considering unitary channels $U$ and adopting the $l_{1}$-norm,
it was proven in \cite{Bu2015,Garcia-Diaz2015} that ${\cal P}_{C}(U)=\widetilde{{\cal P}}_{C}(U)$
in the case of a single qubit, but ${\cal P}_{C}(U)<\widetilde{{\cal P}}_{C}(U)$
strictly in any dimension larger than $2$. These results were then shown to hold for arbitrary non-unitary channels $\Lambda$ in \cite{Bu2016}, meaning that in general the maximum coherence gain due to a qudit channel is obtained
when acting on an input state with already nonzero coherence.

In addition to channels one may also consider evolutions that are
generated by a Hamiltonian $H$ or a Lindbladian ${\cal L}$ \cite{Garcia-Diaz2015}.
For a time evolution $\Phi_{t}=e^{{\cal L}t}$ we determine the coherence
rate
\begin{equation}
\Upsilon({\cal L})=\lim_{\Delta t\rightarrow0}\frac{1}{\Delta t}\max_{\varrho}[C(e^{{\cal L}\Delta t}\varrho)-C(\varrho)]
\end{equation}
and in case of unitary evolutions $U(t)=e^{-iHt}$ we write
\begin{equation}
\Upsilon(H)=\lim_{\Delta t\rightarrow0}\frac{1}{\Delta t}\max_{\varrho}[C(e^{-iH\Delta t}\varrho e^{iH\Delta t})-C(\varrho)].
\end{equation}

Further alternative approaches to define and quantify the cohering power of quantum channels have been pursued recently in \cite{Zanardi2016a,Zanardi2016b}.
Finally, we mention that similar studies have been done in
entanglement theory \cite{Zanardi2000,Linden2009}. In particular,
\cite{Linden2009} showed that entangling and disentangling power
of unitaries are not equivalent in general.

\subsection{Average coherence of random states and typicality}

While some dynamical properties of coherence may be very dependent
on specific channels and initial states, it is also interesting to
study \textit{typical} traits of coherence quantifiers on randomly
sampled pure or mixed states. Note that generic random states, exhibiting
the typical features of coherence summarized in the following, can
be in fact generated by a dynamical model of a quantized deterministic
chaotic system, such as a quantum kicked top \cite{Puchala2015}.

In \cite{Singh2015c} the authors showed that the relative entropy
of coherence (equal to the distillable coherence), the coherence of
formation, and the $l_{1}$-norm of coherence, all exhibit the {\em
concentration of measure phenomenon}, meaning that, with increasing
dimension of the Hilbert space, the overwhelming majority of randomly
sampled pure states have coherence (according to those quantifiers)
taking values very close to the average coherence over the whole Hilbert
space. This was proven rigorously resorting to L\'evy's lemma, hence
showing that states with coherence bounded away from its average value
occur with exponentially small probability.

\subsubsection{Average relative entropy of coherence}

The exact average of the relative entropy of coherence for pure $d$-dimensional
states $\ket{\psi}\in\mathbb{C}^{d}$ sampled according to the Haar
measure was computed in \cite{Singh2015c}, finding
\begin{equation}
\mathbb{E}C_{r}(\ket{\psi})=H_{d}-1\,,\label{eq:AverageCrP}
\end{equation}
where $H_{d}=\sum_{k=1}^{d}(1/k)$ is the $d$th harmonic number.
As shown in \cite{Puchala2015}, for large dimension $d\gg1$ the
average in Eq.~(\ref{eq:AverageCrP}) tends to $\mathbb{E}C_{r}(\ket{\psi})\simeq\log d-(1-\gamma)$,
with $\gamma\approx0.5772$ denoting the Euler constant. This shows
that random pure states have relative entropy of coherence close to
(but strictly smaller than by a constant value) the maximum $\log d$
\cite{Singh2015c,Puchala2015}.

Typicality of the relative
entropy of coherence for random mixed states was investigated in \cite{Zhang2015a,Zhang2016a,Puchala2015}.
Considering the probability
measure induced by partial tracing, that is, corresponding to random
mixed states $\varrho=\mathrm{Tr}_{\mathbb{C}^{d'}}\ket{\psi}\bra{\psi}$,
with $\ket{\psi}\in\mathbb{C}^{d}\otimes\mathbb{C}^{d'}$ ($d \leq d'$) sampled
according to the Haar measure, \cite{Zhang2015a,Zhang2016a} derived a compact analytical formula for the average $\mathbb{E}C_{r}(\varrho)$, given by
$\mathbb{E}C_{r}(\varrho)=(d-1)/(2d')$.
In particular, when considering the flat Hilbert-Schmidt measure (obtained
by setting $d'=d$ in the previous expression), one gets
\begin{equation}
\mathbb{E}C_{r}(\varrho)=\frac{1}{2}-\frac{1}{2d}\,,\label{eq:AverageCrM}
\end{equation}
which tends asymptotically to a constant $1/2$ for $d\rightarrow\infty$.
The expression in Eq.~(\ref{eq:AverageCrM}) was also independently calculated in \cite{Puchala2015}, in the limit of large dimension $d \gg 1$.
The concentration of measure phenomenon for $C_r$ was then proven in \cite{Zhang2015a,Puchala2015}.
These results show that random mixed states have significantly less (relative entropy of) coherence
than random pure states.

\subsubsection{Average $l_{1}$-norm of coherence}

Concerning the $l_{1}$-norm of coherence, \cite{Singh2015c} derived
a bound to the average $\mathbb{E}C_{l_{1}}(\ket{\psi})$ for pure
Haar-distributed $d$-dimensional states ${\ket{\psi}}$, exploiting
a relation between the $l_{1}$-norm of coherence and the so-called
classical purity \cite{Cheng2015}. Then \cite{Puchala2015} obtained
exact results for the average $C_{l_{1}}$ of pure and mixed states
in large dimension $d\gg1$, respectively distributed according to
the Haar and Hilbert-Schmidt measures, finding
\begin{equation}
\mathbb{E}C_{l_{1}}(\ket{\psi})\simeq(d-1)\frac{\pi}{4}\,,\quad\mathbb{E}C_{l_{1}}(\varrho)\simeq\sqrt{d}\frac{\sqrt{\pi}}{2}\,.\label{eq:AverageCl1}
\end{equation}
This shows that, for asymptotically large $d$, the $l_{1}$-norm
of coherence of random pure states scales linearly with $d$ and stays
smaller than the maximal value $(d-1)$ by a factor $\pi/4$, while
the $l_{1}$-norm of coherence of random mixed states scales only
with the square root of $d$.

\subsubsection{Average recoverable coherence}

In \cite{Miatto2015}, the authors considered a qubit interacting
with a $d$-dimensional environment, of which an $a$-dimensional
subset is considered accessible, while the remaining $k$-dimensional
subset (with $d=ak$) is unaccessible. For illustration, one can think
of the environment being constituted by $N$ additional qubits, of
which $N_{A}$ are accessible and $N_{K}=N-N_{A}$ are unaccessible;
in this case $d=2^{N},a=2^{N_{A}},k=2^{N_{k}}$. While for $d\gg1$
such an interaction leads to decoherence of the principal qubit, its
coherence can be partially recovered by quantum erasure, which entails
measuring (part of) the environment in an appropriate basis to erase
the information stored in it about the system, hence restoring coherence
of the latter. The authors considered random pure states $\ket{\psi}\in\mathbb{C}^{2}\otimes\mathbb{C}^{d}$
of the system plus environment composite, and studied the average
recoverable $l_{1}$-norm of coherence $\mathbb{E}C_{l_{1}}(\varrho)$
of the marginal state $\varrho$ of the principal qubit, following
an optimal measurement on the accessible $a$-dimensional subset of
the environment. They found that the average recoverable coherence\footnote{Note that the term ``recoverable coherence'' is here used in a different
context and refers to a different concept than the one introduced
in \cite{Matera2015} and discussed in Sec.~\ref{sub:Recoverable-coherence}.} stays close to zero if $a<k$, scaling as $\mathbb{E}C_{l_{1}}(\varrho)\propto1/\sqrt{k}$,
but it transitions to a value close to unity as soon as at least half
of the environment becomes accessible, scaling linearly as $\mathbb{E}C_{l_{1}}(\varrho)=1-k/(4a)$
for $a\geq k$. With increasing dimension $d$, the transition at
$a=k$ becomes sharper and the distribution of $C_{l_{1}}(\varrho)$
becomes more concentrated near its average value, the latter converging
to $1$ in the limit $d\rightarrow\infty$. By virtue of typicality,
this means that, regardless of how a high-dimensional environment
is partitioned, suitably measuring half of it generically suffices
to project a qubit immersed in such environment onto a near-maximally
coherent state, a fact reminiscent of the quantum Darwinism approach
to decoherence \cite{Zurek2009QD,Brandao2015QD}.

\subsection{Quantum speed limits}

\label{sec:speed}

In the dynamics of a closed or open quantum system, quantum speed
limits dictate the ultimate bounds imposed by quantum mechanics on
the minimal evolution time between two distinguishable states of the
system. In particular, consider a quantum system which evolves, according
to a unitary dynamics generated by a Hamiltonian $H$, from a pure
state $\ket{\psi}$ to a final orthogonal state $\ket{\psi_{\bot}}=e^{-iH\tau_{\bot}/\hbar}\ket{\psi}$
with $\langle\psi_{\bot}|\psi\rangle=0$. Then, seminal investigations
showed that the evolution time $\tau_{\bot}$ is bounded from below
as follows:
\begin{equation}
\tau_{\bot}(\ket{\psi})\geq\max\{\pi\hbar/(2\Delta E),\,\pi\hbar/(2E)\}\,,\label{eq:QSL}
\end{equation}
where $(\Delta E)^{2}=\langle H^{2}\rangle_{\psi}-\langle H\rangle_{\psi}^{2}$
and $E=\langle H\rangle_{\psi}-E_{0}$ (with $E_{0}$ the ground state
energy), and the two bounds in the right-hand side of Eq.~(\ref{eq:QSL})
are due to \cite{MT1945} and \cite{ML1998}, respectively. In the
last seven decades, a great deal of work has been devoted to identifying
more general speed limits, for pure as well as mixed states, and unitary
as well as non-unitary evolutions; see e.g.~\cite{delCampo2013,Taddei2013,Pires2016} and
references therein. Recent studies have shown in particular that quantifiers
of coherence, or more precisely of asymmetry, play a prominent role
in the determination of quantum speed limits.

The authors of \cite{Marvian2015} studied, for any $\epsilon>0$,
the minimum time $\tau_{\epsilon}^{D}$ necessary for a state $\varrho$
to evolve under the Hamiltonian $H$ into a partially distinguishable
state $\varrho_{t}=e^{-iHt}\varrho e^{iHt}$, such that $D(\varrho,\varrho_{t})\geq\epsilon$
according to a distance functional $D$. In formula,

\begin{equation}
\tau_{\epsilon}^{D}(\varrho)=\begin{cases}
\infty, & \mathrm{if}\,D(\varrho,\varrho_{t})<\epsilon\,\forall t>0;\\
\underset{t>0}{\min}\,\{t:D(\varrho,\varrho_{t})\geq\epsilon\}, & \mathrm{otherwise}.
\end{cases}
\end{equation}

Then, for any $\epsilon>0$ and for any distance $D$ which is contractive
and jointly convex, \cite{Marvian2015} proved that $1/\tau_{\epsilon}^{D}(\varrho)$,
which represents the average speed of evolution, is an asymmetry monotone
of $\varrho$ with respect to time translations generated by the Hamiltonian
$H$, being in particular monotonically nonincreasing under the corresponding
TIO. Interestingly, for pure states, even the inverses of the Mandelstam-Tamm
and Margolus-Levitin quantities appearing in the right-hand side of
Eq.~(\ref{eq:QSL}) are themselves asymmetry monotones, bounding
the asymmetry monotone given by $1/\tau_{\bot}(\ket{\psi})$. The
authors of \cite{Marvian2015} then derived new Mandelstam-Tamm--type
quantum speed limits for unitary dynamics based on various measures
of distinguishability, including a bound featuring the Wigner-Yanase
skew information with respect to $H$ (obtained when $D$ is set to
the relative Rényi entropy of order $1/2$), which was also independently
obtained in \cite{Mondal2015}.

The authors of \cite{Pires2016} developed a general approach to Mandelstam-Tamm--type
quantum speed limits for all physical processes. Given a metric $g$
on the quantum state space, let $\ell_{\Lambda}^{g}(\varrho,\varrho_{\tau})$
denote the length of the path connecting an initial state $\varrho$
to a final state $\varrho_{\tau}=\Lambda[\varrho]$ under a (generally
open) dynamics $\Lambda$. Quantum speed limits then ensue from a
simple geometric observation, namely that the geodesic connecting
$\varrho$ to $\varrho_{\tau}$, whose length can be indicated by
$\mathcal{L}^{g}(\varrho,\varrho_{\tau})$, is the path of shortest
length among all physical evolutions between the given initial and
final states: $\mathcal{L}^{g}(\varrho,\varrho_{\tau})\leq\ell_{\Lambda}^{g}(\varrho,\varrho_{\tau})$
$\forall\Lambda$. Then \cite{Pires2016} considered the infinite
family of quantum speed limits derived from this geometric principle,
with $g$ denoting any possible quantum Riemannian contractive metric \cite{Morozova1991,Petz1996,Petz2010},
including two prominent asymmetry monotones: quantum Fisher information
and Wigner-Yanase skew information (see Sec.~\ref{sub:Asymmetry}).
For any given dynamics $\Lambda$ and pair of states $\varrho,\varrho_{\tau}$,
one can identify the tightest speed limit as the one corresponding
to the metric $g$ such that the length of the dynamical path $\ell_{\Lambda}^{g}$
is the closest to the corresponding geodesic length. In formula, the
tightest speed limit is obtained by minimizing the ratio
\begin{equation}
\delta_{\Lambda}^{g}=\frac{\ell_{\Lambda}^{g}(\varrho,\varrho_{\tau})-\mathcal{L}^{g}(\varrho,\varrho_{\tau})}{\mathcal{L}^{g}(\varrho,\varrho_{\tau})}\,,\label{eq:QSLtightness}
\end{equation}
over the metric $g$. Several examples are presented in \cite{Pires2016},
demonstrating the importance of choosing different information metrics
for open system dynamics, as well as clarifying the roles of classical
populations versus quantum coherences in the determination and saturation
of the speed limits. In particular, in the case of a single qubit,
while for any unitary dynamics the speed limit based on the quantum
Fisher information \cite{Taddei2013} is always tighter than the one
based on the Wigner-Yanase skew information, this is no longer true
when considering non-unitary dynamics. Specifically, for parallel
and transversal dephasing, as well as amplitude damping dynamics,
\cite{Pires2016} derived new tighter speed limits based on the Wigner-Yanase
skew information. We finally mention that looser speed limits involving
the skew information have also been recently presented in \cite{Diego15,Mondal2015}.

The speed of a two-qubit photonic system --- quantified by the family of asymmetry monotones associated with the quantum Riemannian contractive metrics --- undergoing a controlled unitary evolution, has been measured experimentally in \cite{Girolami2016}, by means of an all-optical direct detection scheme requiring less measurements than full state tomography.

\section{\label{sec:applications}Applications of quantum coherence}

In this Section we discuss applications of quantum coherence to a variety of fields, ranging from quantum information processing to quantum sensing and metrology, thermodynamics and biology. Particular emphasis will be given to those settings in which a specific coherence monotone introduced in Sec.~\ref{sec:Quantifying-coherence} acquires an operational interpretation, hence resulting in novel insights stemming from the characterization of quantum coherence as a resource.

\subsection{\label{sec:thermodynamics}Quantum thermodynamics}

Recently, the role of coherence in quantum thermodynamics has been
discussed by several authors. We will review the main concepts in
the following, being in large part based on the resource theory of quantum thermodynamics \cite{FGSL13,Gour2013,Goold2016}  defined by the framework of thermal
operations \cite{Janzing00}.

\subsubsection{Thermal operations}

In the following, we consider a system $S$ and an environment $E$
with a total Hamiltonian $H_{SE}=H_{S}+H_{E}$. Given a state of the
system $\varrho^{S}$, a \textit{thermal operation} on this state
is defined as \cite{Janzing00}
\begin{equation}
\Lambda_{\mathrm{th}}\left[\varrho^{S}\right]=\mathrm{Tr}_{E}\left[U\varrho^{S}\otimes\gamma^{E}U^{\dagger}\right],\label{eq:thermal}
\end{equation}
where $\gamma^{E}=e^{-\beta H_{E}}/\mathrm{Tr}[e^{-\beta H_{E}}]$
is a thermal state of the environment with the inverse temperature
$\beta=1/k_{\rm B} T$ and we demand that the unitary $U$ commutes with the
total Hamiltonian: $[U,H_{SE}]=0$.

The importance of thermal operations arises from the fact that they
are consistent with the first and the second law of thermodynamics
\cite{Rudolf214}. In particular, since the unitary $U$ commutes
with the total Hamiltonian, these operations preserve the total energy
of system and environment. Moreover, they do not increase the Helmholtz
free energy\footnote{As customary in thermodynamics, in this section we define the von
Neumann entropy $S(\varrho)=-\mathrm{Tr}[\varrho\log\varrho]$ with
logarithm to base~$e$.}
\begin{equation}
F(\varrho^{S})=\mathrm{Tr}[\varrho^{S}H_{S}]-k_{\rm B} T S(\varrho^{S}),
\end{equation}
i.e., for any two states $\varrho^{S}$ and $\sigma^{S}=\Lambda_{\mathrm{th}}[\varrho^{S}]$
it holds that $F(\sigma^{S})\leq F(\varrho^{S})$.

Thermal operations also have two other important properties \cite{Rudolf114,Rudolf214}.
First, they are TIO with respect to the Hamiltonian $H_{S}$, i.e.,
\begin{eqnarray}
\Lambda_{\mathrm{th}}\left[e^{-iH_{S}t}\varrho^{S}e^{iH_{S}t}\right] & = & e^{-iH_{S}t}\Lambda_{\mathrm{th}}\left[\varrho^{S}\right]e^{iH_{S}t}.
\end{eqnarray}
Second, they preserve the Gibbs state: $\Lambda_{\mathrm{th}}[\gamma^{S}]=\gamma^{S}$.
As discussed in \cite{Rudolf114,Rudolf214}, these two properties
are related to the first and the second law of thermodynamics respectively.
In particular, preservation of the Gibbs state implies that no work
can be extracted from a thermal state.

A closely related concept is known as \emph{Gibbs-preserving operations}
\cite{Ruch1976,Faist2015}. Here, preservation of the thermal state
is the only requirement on the quantum operation. Interestingly, it
was shown by \cite{Faist2015} that these maps are strictly more powerful
than thermal operations. In particular, Gibbs-preserving operations
can create coherence from incoherent states, while this cannot be
done via thermal operations.

\subsubsection{State transformations via thermal operations}

Several recent works studied necessary and sufficient conditions for
two states $\varrho$ and $\sigma$ to be interconvertible via thermal
operations. In the absence of coherence, i.e., if $\varrho^{S}$ is
diagonal in the eigenbasis of $H_{S}$, such conditions were presented
in \cite{Horodecki2013} and termed thermo-majorization\footnote{Note that thermo-majorization is also related to the mixing distance
studied in \cite{Ruch1978}. We also refer to \cite{Egloff2015},
where a relation between majorization and optimal guaranteed work
extraction up to a risk of failure was investigated.}. More general conditions which allow for the addition of ancillas
and catalytic conversions \cite{Jonathan1999a}, known as \emph{second
laws of quantum thermodynamics}, were presented in \cite{Brandao2015a}.
They can also be applied to the situation where the state $\varrho^{S}$
has coherence. Interestingly, for interconversion of single-qubit
states via thermal operations, \cite{Oppenheim14} found a simple
set of necessary and sufficient conditions, in terms of so-called
damping matrix positivity. Similar considerations were also presented
for other classes of operations such as enhanced thermal operations
\cite{Oppenheim14} and cooling maps \cite{Varun14}.

In the above discussion the state of the environment $\gamma^{E}$
was assumed to be a thermal state. In recent literature on quantum
thermodynamics, this constraint has been relaxed, allowing $\gamma^{E}$
to be a general state of the environment. As discussed in \cite{Lostaglio2015},
it is important to distinguish between two cases: namely, whether
the state $\gamma^{E}$ is incoherent, or has nonzero coherence, with
respect to the eigenbasis of $H_{E}$.

In both cases it is also assumed that an arbitrary number of copies
of $\gamma^{E}$ are available, which is a usual assumption from the
point of view of resource theories. An important result in this respect
was obtained in \cite{Rudolf214}. There, it was shown that by allowing
for an arbitrary number of copies of the Gibbs state together with
some other incoherent state $\gamma^{E}$, the operation in Eq.~(\ref{eq:thermal})
can be used to approximate any incoherent state of the system, i.e.,
any state which is diagonal in the eigenbasis of the Hamiltonian $H_{S}$.
Moreover, in this case it is possible to implement any TIO \cite{Rudolf214}.
Although these processes can be used to perform an arbitrary amount
of work, they are still limited in the sense that they cannot create
coherence \cite{Lostaglio2015}. The situation changes if the state
$\gamma^{E}$ has coherence. In this case the process in Eq.~(\ref{eq:thermal})
can also be used to perform an arbitrary amount of work, and apart
from that the process can also create coherence in the system. However,
even in this case the theory is nontrivial, i.e., not all transformations
are possible \cite{Rudolf214,Lostaglio2015}.

The role of coherence in quantum thermodynamics was further studied
in \cite{Misra2016}, where the authors analyzed the physical situation
in which the resource theories of coherence and thermodynamics play
competing roles. In particular, they  investigated  creation of coherence
for a quantum system (with respect to the eigenbasis of its Hamiltonian $H$)
via unitary operations from a thermal state, and also explored the
energy cost for such coherence creation. Given an initial thermal state  $\varrho_T =e^{-\beta H}/\mathrm{Tr}[e^{-\beta H}]$ at temperature $T=1/\beta$ (setting $k_{\rm B}=1$), \cite{Misra2016} showed that there always exists a unitary transformation (in fact, a real orthogonal one) which maps $\varrho_T$ into a state $\varrho'$ such that its diagonal $\Delta[\varrho']=\varrho_{T'}$ amounts to a thermal state at temperature $T' >T$. This creates the maximal relative entropy of coherence $C_{r,\max}^{\Delta E} = S(\varrho_{T'})-S(\varrho_{T})$, at the cost of spending an amount of energy $\Delta E =\mathrm{Tr}[H (\varrho_{T'}-\varrho_{T})]$.

\subsubsection{Work extraction and quantum thermal machines}

Interestingly, coherence cannot be converted to work in a direct way.
This phenomenon is known as \emph{work locking} \cite{Horodecki2013,Skrzypczyk2014,Rudolf114},
and can be formalized as follows \cite{Korzekwa2015}:
\begin{equation}
\left\langle W\right\rangle \left(\varrho^{S}\right)\leq\left\langle W\right\rangle \left(\Pi\big[\varrho^{S}\big]\right).
\end{equation}
Here, $\left\langle W\right\rangle (\varrho^{S})$ denotes the amount
of work that can be extracted from the state $\varrho^{S}$, and $\Pi[\varrho^{S}]=\sum_{i}\mathrm{Tr}\big[\Pi_{i}\varrho^{S}\big]\Pi_{i}$,
where $\Pi_{i}$ are projectors onto the eigenspaces of $H_{S}$.
Note that the operation $\Pi$ is in general different from the full
dephasing $\Delta$ defined in Eq.~(\ref{eq:dephasing}), since the
latter removes all offdiagonal elements, while $\Pi$ preserves some
offdiagonal elements if the Hamiltonian $H_{S}$ has degeneracies.
A detailed study of this problem was also presented in \cite{Korzekwa2015},
where it was shown that work extraction from coherence is still possible
in certain scenarios. This relies on the repeated use of a coherent ancilla in a catalytic way as shown by \cite{Aberg14}.
Further results on the role of coherence for work extraction have also been presented in \cite{Kammerlander2015}.
Moreover, it was shown in \cite{Vacanti2015} that work is typically
required for keeping coherent states out of thermal equilibrium. The role of coherence in determining the distribution of work done on a quantum system has been also studied in \cite{Solinas2015,Solinas2016}.

The role played by coherence in the operation of quantum thermal machines, such as heat engines and refrigerators, has been investigated recently \cite{Rahav2012,Scully2011}.
Various authors have explored the use of optical coherence, in the form of squeezing in a thermal bath, to push the performance of nanoscale heat engines and quantum absorption refrigerators beyond their classical limitations \cite{Abah2014,Lutz2014,Correa2014,David2015,Zambrini2016}. However, the advantages found
in these studies are not directly related to a processing of coherence, but originate at least in part from
the fact that, in energetic terms, a squeezed bath has an energy content which is equivalent to that
of a thermal bath at a higher effective temperature.
Quantum coherence was also shown to be useful for transient cooling in absorption refrigerators \cite{Mitchison2015}. More generally, the authors of \cite{UzdinPRX} established the thermodynamical equivalence of all engine types in the quantum regime of small action (compared to $\hbar$). They then identified generic coherent and incoherent work extraction mechanisms, and showed that coherence enables power outputs that can reach significantly beyond the power of incoherent (i.e., stochastic) engines.

It is noteworthy that the control of any engine, especially an autonomous
device, requires a clock in order to switch on and off an interaction at specified
moments in time, and thereby control the device. At the quantum level, such a control
leads to correlations and thus a possible loss of coherence in the clock. \cite{WoodsSO16}
addresses the question of the coherence cost of such control via clocks and establishes
limits on the backaction on the clock, and therefore its resource consumption, in terms
of energy and coherence.

\subsection{\label{sub:Quantum-algorithms}Quantum algorithms}

The role of coherence in quantum algorithms was discussed by \cite{Hillery2015},
with particular focus on the Deutsch-Jozsa algorithm \cite{Deutsch1992}.
This quantum algorithm can decide whether a boolean function is constant
or balanced by just one evaluation of the function, while in the classical
case the number of evaluations grows exponentially in the number of
input bits. As was shown by \cite{Hillery2015}, coherence is a resource
in this protocol in the sense that a smaller amount of coherence in
the protocol increases the error of guessing whether the underlying
function was constant or balanced.

A similar investigation with respect to the Grover algorithm
\cite{Grover1997} was performed in \cite{Anand2016,Shi2016}.
In \cite{Anand2016}, the authors studied the relation between coherence
and success probability in the analog Grover algorithm, which is a
version of the original Grover algorithm based on adiabatic Hamiltonian
evolution. It was found that the success probability $p_{\text{succ}}$ of the algorithm
is related to the amount of coherence in the corresponding quantum
state as follows \cite{Anand2016}:
\begin{align}
C_{l_{1}}(p_{\text{succ}}) & =2\sqrt{p_{\text{succ}}(1-p_{\text{succ}})},\\
C_{r}(p_{\text{succ}}) & =-p_{\text{succ}}\log_{2}p_{\text{succ}}-(1-p_{\text{succ}})\log_{2}(1-p_{\text{succ}}).
\end{align}

Another important quantum algorithm is known as deterministic quantum
computation with one qubit (DQC1) \cite{Knill1998}. This quantum
algorithm provides an exponential speedup over the best known classical
procedure for estimating the trace of a unitary matrix (given as a
sequence of two-qubit gates). Interestingly, this algorithm requires
vanishingly little entanglement\footnote{In this context we mention that only bipartite entanglement was considered
in \cite{Datta2005}, and the role of multipartite entanglement in
DQC1 remains unclear. We refer to \cite{Parker2000,ParkerP2002} for
similar considerations with respect to Shor's algorithm.} \cite{Datta2005}, but a typical instance of DQC1 has nonzero quantum
discord \cite{Datta2008}. The role of quantum discord for DQC1 was
later questioned by \cite{Dakic2010}, who showed that certain nontrivial
instances do not involve any quantum correlations. This issue was further discussed in  \cite{Datta2011}.
Thus, the question of which type of quantumness correctly captures
the performance of this algorithm remained open. The role of coherence
for DQC1 was first studied by \cite{Ma2015}, and later by \cite{Matera2015}.
The latter work indicates that coherence is indeed a suitable figure
of merit for this protocol. In particular, \cite{Matera2015} showed
that the precision of the algorithm is directly related to the recoverable
coherence, defined in Sec.~\ref{sub:Recoverable-coherence}.

\subsection{Quantum metrology}

\label{sec:metro}

The main goal of quantum metrology \cite{Braunstein1994,Braunstein1996,Giovannetti2004,Giovannetti2006}
is to overcome classical limitations in the precise estimation of
an unknown parameter $\varphi$ encoded e.g.~in a unitary evolution
$U_{\varphi}=e^{-i\varphi H}$. Applications of quantum metrology
include phase estimation for accelerometry, optical and gravitational
wave interferometry, high precision clocks, navigation devices, magnetometry,
thermometry, remote sensing, and superresolution imaging \cite{Paris2009,Giovannetti2011}.

As one can appreciate by the following simple example, quantum coherence
plays a fundamental role in this task. For simplicity, let $H$ be
a nondegenerate single-qubit Hamiltonian $H=E_{0}\ket{0}\bra{0}+E_{1}\ket{1}\bra{1}$.
A very simple possibility to estimate $\varphi$ is to apply the unitary
to a single-qubit state $\ket{\psi}=a\ket{0}+b\ket{1}$, and to perform
a measurement on the final state $U_{\varphi}\ket{\psi}=ae^{-i\varphi E_{0}}\ket{0}+be^{-i\varphi E_{1}}\ket{1}$.
If the probe state $\ket{\psi}$ has no coherence in the eigenbasis
of $H$ (i.e., $a=0$ or $b=0$), the final state $U_{\varphi}\ket{\psi}$
will be the same as $\ket{\psi}$ up to an irrelevant global phase,
i.e., from the final measurement we cannot gain any information about
the parameter $\varphi$. On the other hand, if $a$ and $b$ are
both nonzero, it is always possible to extract information about $\varphi$
via a suitable measurement.

In general, given an initial probe state $\varrho$ and assuming the
probing procedure is repeated $n$ times, the mean square error $(\Delta\varphi)^{2}$
in the estimation of $\varphi$ is bounded below by the quantum Cramér-Rao
bound \cite{Braunstein1994}
\begin{equation}
(\Delta\varphi)^{2}\geq\frac{1}{nI(\varrho,H)},\label{eq:QCR}
\end{equation}
where $I(\varrho,H)$ is the quantum Fisher information, a quantifier
of asymmetry (i.e., of unspeakable coherence) \cite{Marvian2016}
defined in Eq.~(\ref{eq:QFI}). As the bound in Eq.~(\ref{eq:QCR})
is asymptotically achievable for $n\gg1$ by means of a suitable optimal measurement, the quantum Fisher information
directly quantifies the optimal precision of the estimation procedure,
and is thus regarded as the main figure of merit in quantum metrology
\cite{Paris2009,Giovannetti2011}.

Using only probe states without any coherence or entanglement, the
quantum Fisher information can scale at most linearly with $n$, $I(\varrho,H)\sim n$.
However, starting from a probe state $\varrho$ with coherence (e.g.,
the maximally coherent state given by $a=b=1/\sqrt{2}$ in the previous
example) and applying $U_{\varphi}$ sequentially $n$ times before
the final measurement, allows one to reach the so-called Heisenberg
scaling, $I(\varrho,H)\sim n^{2}$, which yields a genuine quantum
enhancement in precision \cite{Giovannetti2006}. In this clear sense,
quantum coherence in the form of asymmetry is the primary resource
behind the power of quantum metrology.

More generally, \cite{Marvian2016} proved that any function which
quantifies the performance of probe states $\varrho$ in the metrological
task of estimating a unitarily encoded parameter $\varphi$ should
be a quantifier of asymmetry with respect to translations $U_{\varphi}$
induced by the generator $H$. Notice that if the parameter $\varphi$
is identified with time $t$, the quantum Fisher information and related
quantifiers of asymmetry \cite{Girolami2016}, as discussed in Sec.~\ref{sub:Asymmetry}, acquire an interpretation as the speed of
evolution of the probe state $\varrho$ under the dynamics $U_{t}$
generated by $H$. This  highlights the role of coherence quantifiers
in the determination of quantum speed limits, as reviewed in Sec.~\ref{sec:speed}.

In absence of noise, the Heisenberg scaling can be equivalently achieved
using $n$ entangled probes in parallel, each subject to one instance
of $U_{\varphi}$ \cite{Huelga97,Giovannetti2006}. A great deal of work has
been devoted to characterizing possibilities and limitations for quantum
metrology in the presence of various sources of noise, which result
in loss of coherence or entanglement of the probes \cite{Huelga97,escher_general_2011,Demkowicz-DobrzanskiNComms2012,chin2012quantum,Maccone2014,Giovannetti2011,Chaves2012,Smirne2016,Nichols2016,Braun2017}.
Typically the Heisenberg scaling is not retained, except under some
error models which allow for the successful implementation of suitable
quantum error correcting procedures \cite{preskill2000quantum,Macchiavello2002,arrad2014increasing,Kessler2014,Dur2014,Kolo2016,unden2016quantum}.

An alternative investigation on the role of coherence in quantum metrology has been carried out in \cite{Allegra2016}, where a relation has been derived between the quantum Fisher information and the second derivative of the relative entropy of coherence, the latter evaluated with respect to the optimal measurement basis in a (unitary or noisy) parameter estimation process.

\subsection{Quantum channel discrimination}

\label{sec:discr} Quantum coherence also plays a direct role in quantum
channel discrimination, a variant of quantum metrology where the task
is not to identify the value of an unknown parameter $\varphi$, but
to distinguish between a set of possible values $\varphi$ can take.
In the case of a binary channel discrimination, in particular deciding
whether a unitary $U_{\varphi}=e^{-i\varphi H}$ is applied or not
to a local probe (i.e., distinguishing between $U_{\varphi}$ and
the identity), the Wigner-Yanase skew information defined in Eq.~(\ref{eq:WYI})
has been linked to the minimum error probability of the discrimination
\cite{Girolami14,Girolami2013,Farace2014}.

More recently, the task of \textit{quantum phase discrimination} has
been studied in \cite{Napoli2016,Piani2016} (see also Fig.~\ref{fig:plan},
left panel). Consider a $d$-dimensional probe and a set of unitary
channels $\{U_{\varphi}\}$ generated by $H=\sum_{i=0}^{d-1}i\ket{i}\!\bra{i}$,
where $\{\ket{i}\}$ sets the reference incoherent basis for the probe
system and $\varphi$ can take any of the $d$ values $\big\{\frac{2\pi k}{d}\big\}_{k=0}^{d-1}$
with uniform probability $1/d$. One such channel acts on the probe,
initialized in a state $\varrho$, and the goal is to guess which
channel instance has occurred (i.e., to identify the correct value
of $\varphi$) with the highest probability of success. Using any
probe state $\sigma\in{\cal I}$ which is incoherent with respect
to the eigenbasis of $H$, no information about $\varphi$ is imprinted
on the state and the probability $p_{\text{succ}}(\sigma)$ of guessing
its correct value is simply given by $1/d$, corresponding to a random
guess. On the other hand, a probe state $\varrho$ with coherence
in the eigenbasis of $H$, accompanied by an optimal measurement at
the output, allows one to achieve a better discrimination, leading
to a higher probability of success $p_{\text{succ}}(\varrho)\geq p_{\text{succ}}(\sigma)$.

The enhancement in the probability of success for this task when exploiting
a coherent state $\varrho$, compared to the use of any incoherent
state $\sigma\in{\cal I}$, is given exactly by the robustness of
coherence of $\varrho$ defined in Eq.~(\ref{eq:Robustness}) \cite{Napoli2016,Piani2016}:
\begin{equation}
\frac{p_{\text{succ}}(\varrho)}{p_{\text{succ}}(\sigma)}=1+R_{C}(\varrho).\label{RobDisc}
\end{equation}
This provides a direct operational interpretation for the robustness
of coherence $R_{C}$ in quantum discrimination tasks. Such an interpretation
can be extended to more general channel discrimination scenarios (i.e.,
with non-uniform prior probabilities, and including non-unitary incoherent
channels) and carries over to the robustness of asymmetry with respect
to arbitrary groups \cite{Piani2016,Napoli2016}.

\subsection{Witnessing quantum correlations}

Recently, several authors tried to find Bell-type inequalities for
various coherence quantifiers. In particular, \cite{Bu2016} considered
quantifiers of coherence of the form $C(X,Y,\varrho^{AB})$, where
$X$ and $Y$ are local observables on the subsystem $A$ and $B$
respectively, and the coherence is considered with respect to the
eigenbasis of $X\otimes Y$. They found a Bell-type bound for this
quantity for all product states $\varrho^{A}\otimes\varrho^{B}$,
and showed that the bound is violated for maximally entangled states
and a certain choice of observables $X$ and $Y$. In a similar spirit,
the interplay between coherence and quantum steering was investigated
in \cite{Mondal2015a,Mondal2015b}, where steering inequalities for
various coherence quantifiers were found, and in \cite{Hu2015,Hu2015a},
where the maximal coherence of steered states was investigated.

As was further shown by \cite{Girolami2015}, detection of coherence
can also be used to witness multipartite entanglement. In particular,
an experimentally accessible lower bound on the quantum Fisher information
(which does not require full state tomography) can serve as a witness
for multipartite entanglement, as was explicitly demonstrated for
mixtures of GHZ states \cite{Girolami2015}. This builds on previous
results on detecting different classes of multipartite entanglement
using the quantum Fisher information \cite{smerzi_2009,smerzi_2012,toth_2012,smerzi_review_2014,toth_review_2014}.

\subsection{Quantum biology and transport phenomena}

Transport is fundamental to a wide range of phenomena in the natural
sciences and it has long been appreciated that coherence can play
an important role for transport e.g.~in the solid state \cite{solidcoherence,Li2012}.
Recently, however, some research efforts have started to investigate
the role of coherence in the perhaps surprising arena of ``warm,
wet and noisy'' biological systems. Motivated in part by experimental
observations using ultrafast electronic spectroscopy of light-harvesting
complexes in photosynthesis \cite{Engel2007,Collini2010} the beneficial
interplay of coherent and incoherent dynamics has been identified
as a key theme \cite{Plenio2008,Mohseni2008,Caruso2009,Aspuru2009,Rebentrost2009}
in biological transport and more generally in the context of biological
function \cite{Huelga13}. It is now recognized that typically both
coherent {\em and} noise dynamics are required to achieve optimal
performance.

A range of mechanisms to support this claim and understand its origin
qualitatively (see \cite{Huelga13} for an overview) have been identified.
These include constructive and destructive interference due to coherence
and its suppression by decoherence \cite{Caruso2009} as well as the
interaction between electronic and long-lived vibrational degrees
of freedom (coherent) \cite{chin2010noise,chin2013role,kolli2012fundamental,christensson2012origin,o2014non,prior2010efficient,womickmoran2012,roden2016long}
in the environment and with a broadband vibrational background (incoherent)
\cite{Plenio2008,Mohseni2008,Caruso2009,Aspuru2009,Rebentrost2009}.
Nevertheless, the detailed role played by coherence and coherent dynamics
in these settings remains to be unraveled and quantified and it is
here where the detailed quantitative understanding of coherence emerging
from its resource theory development may be beneficial.

Initially, researchers studied the entanglement properties of states
\cite{Caruso2009,sarovar2010quantum,fassioli2010distribution} and
evolutions \cite{caruso2010entanglement} that emerge in biological
transport dynamics. It should be noted though that in the regime of
application the quantities considered by \cite{sarovar2010quantum}
amount to coherence quantifiers rather than purely entanglement quantifiers.
In the studies of the impact of coherence on transport dynamics, formal
approaches using coherence and asymmetry quantifiers based
on the Wigner-Yanase skew information were used \cite{Vatasescu,Vatasescu2},
but the connection to function has remained tenuous so far. It was
indeed noted that it may become necessary to separately quantify real
and imaginary part of coherence as these can have significantly different
effects on transport \cite{roden2016probability}. Another question
of interest in this context concerns that of the distinction between
classical and quantum coherence \cite{o2014non} and dynamics \cite{wilde2010could,Li2012}
in biological systems, most notably photosynthetic units.

Finally, it should be noted that there are other biological phenomena
that are suspected to benefit from coherent and incoherent dynamics,
notably magnetoreception in birds \cite{Ritz2000,Gauger2011,cai2010quantum}
where the role of coherence in the proposed radical pair mechanism
was studied on the basis of coherence quantifiers \cite{kominis2015radical}
and the molecular mechanisms underlying olfaction \cite{turin1996spectroscopic}.
Unlike photosynthesis, however, experimental evidence is still limited
and not yet at a stage where conclusions drawn from the quantitative
theory of coherence as a resource can be verified.

\subsection{Quantum phase transitions}

Coherence and asymmetry quantifiers have been employed to
detect and characterize quantum phase transitions, i.e., changes in
the ground state of many-body systems occurring at or near zero temperature
and driven purely by quantum fluctuations. The critical points can
be identified by witnessing a particular feature in a chosen coherence
quantifier, such as a divergence, a cusp, an inflexion, or a vanishing
point.

The authors of \cite{Karpat2014,Cakmak2015} showed that single-spin
coherence reliably identifies the second-order quantum phase transition
in the thermal ground state of the anisotropic spin-$\frac{1}{2}$
XY chain in a transverse magnetic field. In particular, the single-spin
skew information with respect to the Pauli spin-$x$ operator $\sigma_{x}$,
as well as its experimentally friendly lower bound which can be measured
without state tomography \cite{Girolami14}, exhibit a divergence
in their derivative at the critical point, even at relatively high
temperatures.

The authors of \cite{Malvezzi2016} extended the previous analysis
to ground states of spin-$1$ Heisenberg chains. Focusing on the one-dimensional
XXZ model, they found that no coherence and asymmetry quantifier
(encompassing skew information, relative entropy, and $l_{1}$-norm)
is able to detect the triple point of the infinite-order Kosterlitz-Thouless
transition, while the single-spin skew information with respect to
a pair of complementary observables $\sigma_{x}$ and $\sigma_{z}$
can instead be employed to successfully identify both the Ising-like
second-order phase transition and the SU(2) symmetry point.

Further applications of coherence and asymmetry quantifiers
to detecting quantum phase transitions in fermionic and spin models
have been reported in \cite{LiSrep2016,ChenJJ2016}.

\section{\label{sec:conclusions}Conclusions}

In this Colloquium we have seen that quantum coherence plays an important
role in quantum information theory, quantum thermodynamics, and quantum
biology, as well as physics more widely. Similar to entanglement,
but even more fundamental, coherence can be regarded as a {\it resource},
if the experimenter is limited to quantum operations which cannot
create coherence. The latter set of operations is not uniquely specified: in
Section~\ref{sec:resource-theories} we have
reviewed the main approaches in this direction, their motivation,
and their main differences. Most of these approaches have some desirable
properties which distinguish them from the other frameworks.

It is instrumental to compare once again the resource theory of coherence
to the resource theory of entanglement. The latter has a natural approach
which is defined by the set of LOCC operations. This set of operations
has a clear physical motivation and several nice properties which
make exact evaluation tractable in many relevant situations. In particular,
this approach has a golden unit, since any bipartite quantum state
can be created via LOCC operations from a maximally entangled state.

In the following, we provide six simple conditions which we believe
any physical theory of quantum coherence as a resource should be tested
on. These conditions are motivated by recent developments on the resource
theories of coherence and entanglement. In particular, we propose
that any resource theory of coherence should have a set of free operations
$\mathcal{F}$ with the following properties:
\begin{enumerate}
\item Physical motivation: the set of operations $\mathcal{F}$ has a well
defined physical justification.
\item Post-selection: The set $\mathcal{F}$ allows for post-selection,
i.e., there is a well-defined prescription for performing multi-outcome
measurements, and obtaining corresponding probabilities and post-measurement
states.
\item No coherence creation: The set $\mathcal{F}$ (including post-selection)
cannot create coherence from incoherent states.
\item Free incoherent states: The set $\mathcal{F}$ allows to create any
incoherent state from any other state.
\item Golden unit: The set $\mathcal{F}$ allows to convert the maximally
coherent state $\ket{\Psi_{d}}$ to any other state of the same dimension.
\item No bound coherence: Given many copies of some (coherent) state $\varrho$,
the set $\mathcal{F}$ allows to extract maximally coherent single-qubit
states $\ket{\Psi_{2}}$ at nonzero rate.
\end{enumerate}
While conditions 2--6 can be tested directly, the first condition
seems to be the most demanding. In Table \ref{tab:operations} we
list the status of conditions 2--6 for the existing sets MIO, IO,
SIO, DIO, TIO, PIO, GIO, and FIO, introduced in Sec.~\ref{sub:Classes}.
In place of the first condition, we give the corresponding literature
reference, where suitable motivations can be found for each set. As
the Table shows, several frameworks of coherence do not fulfill all
of our criteria, and several entries still remain open.

\begin{table}
\begin{tabular}{c>{\centering}p{4.1cm}ccccc}
\hline
 & 1  & 2  & 3  & 4  & 5  & 6\tabularnewline
\hline
MIO  & \cite{Aberg2006}  & yes  & yes  & yes  & yes  & yes\tabularnewline
IO  & \cite{Baumgratz2014,Winter2015}  & yes  & yes  & yes  & yes  & yes\tabularnewline
SIO  & \cite{Winter2015,Yadin2015b}  & yes  & yes  & yes  & yes  & ?\tabularnewline
DIO  & \cite{Chitambar2016,Marvian2016}  & yes  & yes  & yes  & yes  & ?\tabularnewline
TIO  & \cite{Marvian2015,Marvian2016}  & yes  & yes  & yes  & no  & ?\tabularnewline
PIO  & \cite{Chitambar2016}  & yes  & yes  & yes  & no  & ?\tabularnewline
GIO  & \multirow{2}{4.1cm}{\centering{}\cite{Streltsov2015d}} & yes  & yes  & no  & no  & no\tabularnewline
FIO  &  & yes  & yes  & no  & no  & ?\tabularnewline
\hline
\end{tabular}

\caption{\label{tab:operations} List of alternative frameworks of coherence
with respect to our criteria 1--6 provided in the text. For the first
criterion we give the corresponding literature reference. Unknown
entries are denoted by ``?''.}
\end{table}

We further reviewed in Section~\ref{sec:Quantifying-coherence}
the current progress on quantifying coherence and related manifestations of nonclassicality in compliance with the underlying framework of resource theories, in particular highlighting interconnections between different measures and, where possible, their relations to entanglement measures. Several open questions remain to be addressed in these topics, as we pointed out throughout the text. The rest of the Colloquium (Sections~\ref{sec:dynamics} and \ref{sec:applications}) was dedicated to investigate more physical aspects of coherence in quantum systems and its applications to quantum technologies, many-body physics, biological transport, and thermodynamics. Most of these advances are still at a very early stage, and the operational value of coherence still needs to be pinpointed clearly in many contexts.

We expect that substantial future research will focus on various aspects
of coherence in physics, information theory, biology, and other branches of science and engineering.
To highlight the ultimate role of quantum coherence
as a resource in these and related research fields, we need to reveal
new phenomena which can be explained in quantitative terms by the presence of coherence,
but cannot be traced back to entanglement, or any other kind of nonclassical resource.
We hope that this Colloquium may pave the way towards further breakthroughs
in this exciting research direction.

\begin{acknowledgments}
We thank Tillmann Baumgratz, Manabendra Nath Bera, Paul Boes, Thomas Bromley, Dagmar
Bru\ss, Kaifeng Bu, Eric Chitambar, Marco Cianciaruso, Marcus Cramer, Animesh Datta,
Maria Garc{\'{i}}a-D{\'{i}}az, Dario Egloff, Jens Eisert, Heng Fan, Ir\'en\'ee Fr\'erot,
Paolo Giorda, Davide Girolami, Gilad Gour, Micha\l{} Horodecki, Pawe\l{} Horodecki, Susana
Huelga, Hermann Kampermann, Nathan Killoran, Maciej Lewenstein, Iman Marvian, Mauricio Matera,
Marco Piani, Swapan Rana, Alexey Rastegin, Bartosz Regula, Tommaso Roscilde, Isabela Almeida
Silva, Paolo Solinas, Robert Spekkens, Thomas Theurer, Werner Vogel, Henrik Wilming, Andreas
Winter, Dong Yang, Lin Zhang, and Karol \.{Z}yczkowski for discussions and feedback on the manuscript.
We acknowledge financial support by the European Research Council (StG GQCOP Grant No.~637352
and SyG BioQ Grant No.~319130), the EU projects PAPETS and QUCHIP, the Foundational Questions
Institute (Grant No.~FQXi-RFP-1601), the National Science Center in Poland (POLONEZ UMO-2016/21/P/ST2/04054), the Alexander von Humboldt-Foundation, the MINECO (SEVERO OCHOA Grant SEV-2015-0522, FISICATEAMO FIS2016-79508-P), the Generalitat de Catalunya (SGR 874 and CERCA Program), and Fundaci\'o Privada Cellex. This publication
was  made possible through the support of a grant from the John Templeton Foundation. The
opinions expressed in this publication are those of the authors and do not necessarily reflect
the views of the John Templeton Foundation.
\end{acknowledgments}

% \bibliographystyle{apsrmp4-1}
%\bibliography{literature}

%merlin.mbs apsrmp4-1.bst 2010-07-25 4.21a (PWD, AO, DPC) hacked
%Control: author (75) reversed first initials jnrlst

\end{document}